\documentclass{aa}  
\usepackage{graphicx}
\usepackage{txfonts}
\usepackage{natbib}  
\usepackage{subfig}

\begin{document}
   \title{An interferometric study of the post-AGB binary 89\,Herculis\thanks{Based on observations made with ESO Telescopes at the La Silla Paranal Observatory under program ID 077.D-0071.}}
   \subtitle{II Radiative transfer models of the circumbinary disk}
   \author{M. Hillen\inst{1}
          \and
          J. Menu\inst{1}
          \and
          H. Van Winckel\inst{1} 
          \and
          M. Min\inst{2}
          \and
          C. Gielen\inst{1,3}
          \and
          T. Wevers\inst{1,4}
          \and
          G. D. Mulders\inst{5}
          \and
          S. Regibo\inst{1}
          \and
          T. Verhoelst\inst{1,3}
          }

   \institute{Instituut voor Sterrenkunde (IvS), KU Leuven,
              Celestijnenlaan 200D, B-3001 Leuven, Belgium\\
              \email{Michel.Hillen@ster.kuleuven.be}
            \and
               Sterrenkundig Instituut Anton Pannekoek, University of Amsterdam, Science Park 904, 1098 XH Amsterdam, The Netherlands
            \and
               Belgian Institute for Space Aeronomy, Brussels, Belgium         
            \and 
               Department of Astrophysics/IMAPP, Radboud University Nijmegen, P. O. Box 9010, 6500 GL Nijmegen, The Netherlands
            \and
               Lunar and Planetary Laboratory, The University of Arizona, 1629 E. University Blvd., Tucson AZ 85721, USA
             }

   \date{Received 03 March, 2014; accepted 07 May, 2014}
   \authorrunning{Hillen et al.}
   \titlerunning{MCMax models of 89 Herculis}

  \abstract
   {The presence of stable disks around post-Asymptotic Giant Branch (post-AGB) binaries is a widespread phenomenon. Also, the presence of (molecular)
   outflows is now commonly inferred in these systems.}
   {In the first paper of this series, a surprisingly large fraction of optical light was found to be resolved in the 89 Her post-AGB binary system. 
   The data showed that this flux arises from close to the central binary. Scattering off the inner rim of the circumbinary disk, 
   or scattering in a dusty outflow were suggested as two possible origins. With detailed dust radiative 
   transfer models of the circumbinary disk, we aim to discriminate between the two proposed configurations.}
   {By including Herschel/SPIRE photometry, we extend the spectral energy distribution (SED) such that it now fully covers UV to sub-mm wavelengths.
   The MCMax Monte Carlo radiative transfer code is used to create a large grid of disk models. Our models include a self-consistent treatment
   of dust settling as well as of scattering. A Si-rich composition with two additional opacity sources, metallic Fe or amorphous C, are tested. 
   The SED is fit together with archival mid-IR (MIDI) visibilities, and the optical and near-IR visibilities of Paper I. In this way we constrain 
   the structure of the disk, with a focus on its inner rim.}
   {The near-IR visibility data require a smooth inner rim, here obtained with a double power-law parameterization of the radial 
   surface density distribution. A model can be found that fits all of the IR photometric and interferometric data well, with either of the two continuum 
   opacity sources. Our best-fit passive models are characterized by a significant amount of $\sim$mm-sized grains, which are settled to the midplane of 
   the disk. Not a single disk model fits our data at optical wavelengths because of the opposing constraints imposed by 
   the optical and near-IR interferometric data. }
   {A geometry in which a passive, dusty, and puffed-up circumbinary disk is present, can reproduce all of the IR, but not the optical observations of 
   89 Her. Another dusty component (an outflow or halo) therefore needs to be added to the system.}


   \maketitle
%

\section{Introduction}
In \citet{2013AAHillen}, hereafter Paper I, we reported the first detection of a large amount (35\% of the total) of optical scattered 
light emerging from the close circumstellar environment of the post-Asymptotic Giant Branch (post-AGB) binary 89 Herculis. Many post-AGB objects display large-scale
structures in scattered light \citep[][]{2010ApJSanchez,2008ApJSiodmiak,2002AABujarrabal,2000ApJUeta}, but generally the post-AGB 
stars in binary systems do not \citep[and are therefore classified as ``stellar'' in e.g.,][]{2008ApJSiodmiak}. These post-AGB binaries 
(with P$\sim$100-3000\,d) do have a circumstellar dust reservoir, but of relatively low mass, in the form of a stable Keplerian circumbinary 
disk \citep[see, e.g., the census of Galactic sources by][]{2006AAdeRuyter}. This disk is presumably very similar in structure to protoplanetary 
disks found around young stars, despite the different formation mechanism. In contrast, post-AGB stars that have pronounced 
aspherical nebulosities, often accompanied by an overdense, dusty equatorial waist or torus that obscures the central 
object \citep[see, e.g., the Hubble Space Telescope (HST) survey of][]{2007AJSahai}, are dubbed protoplanetary nebulae (PPNe). Few binaries, 
if at all, are found at the heart of PPNe \citep{2011ApJHrivnak}.

A notable case within the binary group is the Red Rectangle (RR), which has been imaged in various optical filters, showing an 
extended nebula with a complex spatial as well as chemical structure \citep{2006ApJVijh,2005ApJVijh,2004ApJVijh,2004AJCohen}. 
The RR is also the only post-AGB binary for which the circumbinary disk has been directly imaged 
\citep[][and references therein]{2011MNRASLagadec,2004AJCohen,2002AATuthill,1993ApJBregman}, and for which the Keplerian velocity field 
has been directly resolved in rotational CO lines \citep{2005AABujarrabal}. The recent Atacama Large Millimeter/submillimeter Array (ALMA) 
maps of \citet{2013AABujarrabalC} also show a clear molecular component that coincides with the bipolar outflow seen in optical light. 

A molecular outflow, in the form of an extended hourglass-like structure, was also detected in the 89 Her system by \citet{2007AABujarrabal}.
Recently, \citet{2013AABujarrabalB} presented evidence for the presence of low-velocity molecular outflows in a large sample of 
post-AGB binaries from the shape of the wings in single-dish rotational CO lines. Bipolar outflows might thus be a common property of post-AGB binaries
with disks. But it is unlikely that all of these outflows show the complexity of the RR. There is no evidence for a mixed C- and O-based chemistry 
in most objects as is the case for the RR \citep{2006ApJVijh,1998NaturWaters}. The fact that no outflows have yet been detected in optical light in other
objects may also be due to a lack of spatial resolution in previous resolved observations \citep[e.g., those of][]{2008ApJSiodmiak}. 

Many questions remain about these outflows. How and from where in these complex systems are they driven? 
There are now detections of jetlike outflows in several post-AGB binaries, thanks to recent and extensive spectroscopic 
monitoring campaigns \citep[][]{2013MNRASThomas,2012AAGorlova,2011MNRASThomas,2009ApJWitt}, which suggest an origin in the neighborhood of the 
companion star. A compact continuum-scattered-light component could go unnoticed in these spatially-integrated spectra, 
depending on the inclination of the system. 

Given the pole-on viewing angle (i$\sim$12$^\circ$), it is possible that our detected 35\% optical scattered 
light around 89 Herculis in Paper I, is largely arising in an outflow. Alternatively, the scattered light 
could be fully originating from the inner rim of the circumbinary dust disk. The main argument in favor of the latter
interpretation is the conformance of the sizes at optical and near-IR wavelengths, as derived from the interferometric data presented in Paper I. 
The baseline coverage of these data is insufficient to directly distinguish between these two proposed origins: 1) 
the flux is scattered into our line of sight over an extended surface area of which the outer radius equals the inner rim of the disk; 
in this case the material is most likely positioned above the orbital plane (i.e., in an outflow or a halo) as the dust would otherwise sublimate; 
or 2) it arises from a very thin ring, corresponding to the inner rim of the dust disk as resolved in the near-IR. In any case, 
the observations presented in Paper I require that the projected surface area is more compact at optical than at near-IR wavelengths, 
so the bulk disk surface can be ruled out as a dominant scattering source.

One way to differentiate between these options is to model the energetics of the disk in detail, 
taking all spatial constraints into account. If the disk is responsible for \textit{all} of the circumstellar optical and IR flux, 
then it must have a significantly larger scale height than if an additional component is present in the system. 
With an inclination of only $\sim$12$^\circ$, our line of sight to the disk is moreover unfavorable, given 
the preference of grains to scatter in the forward direction \citep[see also][]{2013AAMulders}. Based on these 
simple arguments, the presence of an outflow in addition to the disk seems natural. An exploration of full radiative transfer 
disk models is still needed, however, to rule out the possibility that a disk alone is capable of explaining the data. 

The goal of this paper is therefore to investigate whether a self-consistent radiative transfer model of the circumbinary 
disk in the 89 Herculis system can be created that reproduces all our observations. First, we present some complementary archival 
mid-IR interferometric and new far-IR photometric observations in Sect.~\ref{section:extraobservations}. Subsequently,
the modeling strategy is described in Sect.~\ref{section:MCMax}, and a first confrontation with the observations is done in Sect.~\ref{section:firstresults}. 
In Sects.~\ref{sect:IR},~\ref{sect:carbon}, and~\ref{sect:optical}, some improvements are made to the model, and the fit to the near-IR, mid-IR 
and optical interferometric data is examined in detail. Finally, in Sects.~\ref{sect:discussion} and~\ref{sect:conclusions} we discuss 
our results and summarize our conclusions.





\section{Observations} \label{section:extraobservations}
A detailed account of our optical and near-IR interferometric data was given in Paper I. Here, we add some archival 
data from the MID-infrared Interferometric instrument (MIDI) on the Very Large Telescope Interferometer (VLTI) 
to the existing data set, as well as new Spectral and Photometric Imaging Receiver instrument (SPIRE) photometry from the Herschel satellite.

\subsection{Additional interferometry} \label{subsection:MIDIdata}
Three VLTI/MIDI observations of 89\,Her, procured in April 2006 and already discussed in \citet{2007AABujarrabal}, were reduced 
with version 2.0 of the EWS software \citep{2004SPIEJaffe}\footnote{\texttt{http://home.strw.leidenuniv.nl/$\sim$jaffe/ews/index.html}}.
This software estimates the MIDI visibilities in a coherent and linear way, based on a shift-and-add algorithm 
applied on individual interferometric frames. Two observations used Auxiliary Telescope (AT) stations D0 and G0, and 
the other observation used stations E0 and G0. The projected baseline angles are 75-90$^\circ$ east of north, and the lengths are 
$\sim$30 and 15\,m.

In the calibration, the raw target visibility $V_\mathrm{sci}$ is divided by the 
instrumental visibility $V_\mathrm{ins}$ for each wavelength:
\begin{equation}
V=\frac{V_\mathrm{sci}}{V_\mathrm{ins}}=\frac{C_\mathrm{sci}}{P_\mathrm{sci}}\,\frac{P_\mathrm{cal}}{C_\mathrm{cal}} V_\mathrm{theo-cal}.\label{eq:cal}
\end{equation}
Here, $C$ and $P$ are the measured correlated flux and photometric counts for the target (sci) and 
calibrator (cal), and $V_\mathrm{theo-cal}$ is the expected calibrator visibility.
This assumes that the calibrator and target observations are performed under the same instrumental conditions. 
For a robust calibration, we interpolate the $V_\mathrm{ins}$ of the 
calibrator directly preceding and succeeding the target observation. To account for the longer term variability of the transfer 
function, we add the standard deviation of $V_\mathrm{ins}$ for all calibrators within four hours of the 
science measurement in quadrature to the visibility errors. On the worst night, the variability of the transfer function is $\sim$20\,\%,
which can (partly) be attributed to the difficulty in estimating 
the photometric counts, given the strong background radiation of the delay-line system. This noise source might be source-dependent
because of the adjustments in the delay lines upon a change in the coordinates on the sky. 

The calibrated visibilities are shown in Fig.~\ref{figure:MIDIvis}. 
The 30\,m baseline observations are partly resolved at the red 
edge of the band, with visibilities of $V\sim0.8$. At the blue edge, the situation is less clear due to the larger errors, but 
both 30\,m observations seem to be in agreement with a slightly resolved object. The single 15\,m baseline observation 
shows a flat visibility curve with $V\sim0.8$. This suggests that 89\,Her is at least as resolved on the 15\,m 
baseline as on the 30\,m baseline, however, we are cautious in our interpretation because the observing conditions were poorer
during the night of the 15\,m observation than for the other two nights (with an average atmospheric coherence time of 2.5 rather than 6 and 11\,ms).
The resulting stronger optical path difference fluctuations make it more difficult to avoid decorrelation effects when reducing 
the data\footnote{This is related to the intrinsic reduction algorithm that aligns the individual frames \citep[see, e.g.,][]{2012SPIEBurtscher}.}, 
which may effectively lower the apparent correlated flux. 
Given the stability of the transfer function during the night, 
we do not exclude the 15-m baseline, but keep in mind that the observation is potentially affected by the poorer atmospheric conditions.

\subsection{PTI fluxes}
In addition to visibilities, fluxes can be extracted from the raw data of the Palomar Testbed Interferometer \citep[PTI,][]{1999ApJColavita}. 
We developed a calibration procedure for these fluxes, which has been detailed 
in \citet{2012AAHillen}. The PTI measured visibilities in five channels within the K band and four channels within the H band. Only 
the central channels within the H and K bands are used in our interferometric analysis, but we include all channels for the fluxcalibration.
As explained in \citet{2012AAHillen}, we separate the calibration into an absolute calibration in one of the channels within each band (at 1.7~$\mu$m
and at 2.2~$\mu$m, where the atmospheric transmission is the highest) and a separate calibration of the spectral shape over the other channels in the 
bands. This nonstandard approach was developed since the spectral shape calibration can be done more accurately, even in less optimal 
atmospheric conditions, and lead to accurate light curves of Mira variables \citep[see][]{2012AAHillen}. For 89 Herculis 
the calibrated fluxes are also stable in time and are averaged to obtain the final fluxes that are shown in Fig.~\ref{figure:PTIfluxes}. 
These data will serve as an independent check of the shape of the SED in the near-IR spectral range and how the best 
models perform at these wavelengths.

\subsection{SPIRE photometry}
The object 89 Her was observed with the SPIRE instrument
\citep[for more info on its scientific capabilities, observing modes, data reduction, calibration, 
and performance, see][]{SPIREGriffin,SPIRESwinyard} aboard the Herschel satellite \citep{2010AAPilbratt} in March 2012 with a 169\,s exposure time 
(Herschel Observation ID 1342240026). The observation was performed simultaneously in three passbands (centered at 250\,$\mu$m, 350\,$\mu$m, and 500\,$\mu$m),
covering an area of $8' \times 8'$ homogeneously. The \textit{naiveMapper} task was used to project each observed \textit{sample} onto a single pixel
in the reconstructed map. The pixel size is $4''$, $6''$, and $9''$ in the three bands, respectively, 
and the corresponding beam FWHM are $18.2''$, $24.9''$, and $36.3''$, respectively. Aperture photometry was used to extract the final fluxes, after sky background subtraction.
The latter was estimated with the standard \textit{daophot} algorithm. The errors are dominated by the absolute flux calibration and were assumed to 
be 15\% \citep{SPIRESwinyard}. The final calibrated fluxes are shown with the rest of the 
SED, in Fig.~\ref{figure:firstbestSED}.

\subsection{Data selection}
Our photometric data were already discussed in Paper I, with the exception of the Plateau de Bure Interferometer (PdBI) 1.3 and 3~mm continuum fluxes, 
which we have taken from \citep{2007AABujarrabal}. 
We decided to exclude the Infra-Red Astronomy Satellite fluxes \citep[IRAS,][]{1984ApJNeugebauer} from our $\chi^2$ calculation, as we found
that they are systematically higher than the AKARI \citep{2007PASJMurakami}, Wide-field Infrared Survey Explorer \citep[WISE,][]{2010AJWright}, SPIRE and 
Submillimetre Common-User Bolometer Array \citep[SCUBA][]{1999MNRASHolland} measurements at similar or longer wavelengths. Since the Infrared Space Observabory (ISO) spectrum is 
very noisy beyond 25\,$\mu$m, as well as to simplify the analysis, we decided to only include it by integrating it over several well-chosen photometric 
passbands in the near- to mid-IR domain. These passbands are chosen so as to fill up the voids in the spectral coverage 
of the real photometry, and are the Infrared Array Camera bands \citep[IRAC, aboard the Spitzer Space Telescope,][]{2005PASPReach} at 5.8 
and 8.0\,$\mu$m, as well as the WISE W2 ($\sim$5\,$\mu$m) and W3 ($\sim$12~$\mu$m) bands, and the Midcourse Space Experiment 
\citep[MSX,][]{1994JSpRoMill} D band at 15\,$\mu$m. Since 
the error on the absolute calibration of the ISO spectrum is of the order of 10\%, this error value is also adopted for the photometry derived from it.

\section{MCMax modeling} \label{section:MCMax}
\subsection{Introducing MCMax}
The 2D Monte-Carlo radiative-transfer code MCMax was used to solve the structure of the disk and subsequently calculate model SEDs and 
visibilities. The MCMax code has been successfully used to model both spatially and spectrally resolved observations of 
different kinds of gas-rich disks \citep{2011AAMulders,2011AAVerhoeff,2013AAAcke}. Our modeling approach was basically the same as presented in 
\citet{2012AAMulders} and we refer to this paper and \citet{2009AAMin} for a full account of the procedures and physics implemented in MCMax. 
In short, MCMax combines the immediate re-emission procedure of \citet{2001ApJBjorkman} with the method of continuous 
absorption by \citet{1999AALucy} to compute the radiative equilibrium temperature stratification throughout the disk. Subsequently, 
the vertical scale height of the disk is self-consistently determined by solving the equations of vertical hydrostatic equilibrium, 
implicitly assuming thermal coupling between the gas and dust, but dynamical decoupling, i.e., dust settling. The final 
disk structure is found by iterating both steps until convergence is reached, 
typically in less than five iterations. Scattering by dust can be treated with a full angle-dependent Mueller matrix for each scattering event, 
or by assuming isotropic scattering, which needs less computation time \citep{2009AAMin}.

All observables are calculated by analytically integrating the computed source function by ray-tracing. 
This includes the visibility: the Fourier transform is analytically calculated on a circularly sampled image. 
The advantage of this approach is that the inner parts of the image are sampled at a higher resolution so that 
visibilities can be reliably computed up to high spatial frequencies. 

\subsection{The model setup} \label{subsection:setup}
We define a ``standard'' radiative-transfer disk model based on some general characteristics for the sample of disks around post-AGB binaries:
\begin{itemize}
 \item \citet{2007AADeroo} made a disk model of IRAS08544-4431 by fitting the total disk mass and inner rim radius to the SED, while fixing all other
 parameters to typical values for the group of binary post-AGB objects or based on other external constraints (e.g., the stellar 
 T$_{\mathrm{eff}}$ from optical spectroscopy). By tuning the inclination and position angle, this ``standard disk model'' 
 reproduced all of their interferometric observables surprisingly well.
 \item All disks around post-AGB binaries (with a few clear exceptions) have an O-rich mineralogy, their mid-IR spectral features 
 are dominated by amorphous and crystalline silicates and their dust typically has a high degree of grain processing \citep{2008AAGielen,2011AAGielen},
 \item All objects for which long-wavelength fluxes are available in the sample study of \citet{2006AAdeRuyter} have blackbody slopes that are indicative 
 of the presence of a component of mm-sized grains.
 \item \citet{2011ApJSahai} presents evidence for the presence of grains $>$1\,mm in the compact disks of RV Tau, U Mon, and AC Her, from 
 flux measurements at sub-mm and mm wavelengths. The derived dust masses lie between 5\,10$^{-4}$ and 1\,10$^{-3}$\,M$_\odot$.
\end{itemize}

The overall shape of the SED of 89 Her is very reminiscent \citep[see also][]{2006AAdeRuyter} 
of a group 2 Herbig star as defined by \citet{2001AAMeeus}. We therefore include dust settling in our disk model in a 
self-consistent manner \citep[see][for more details about the implementation and consequences 
of this approach]{2012AAMulders}. Our dust composition and grain shape are the same as in the latter paper \citep[i.e., irregular shapes, approximated 
by the distribution of hollow spheres (DHS)][]{2005AAMin}, except that we start with metallic Fe as the additional featureless ``continuum'' opacity source 
($\sim$15\% in mass fraction) instead of amorphous carbon, which would be more difficult to reconcile with the oxygen-rich circumbinary environment. 
The majority of the dust mass consists of silicates \citep[for the composition see]{2007AAMin}.

The dust is distributed over a range of grain sizes with a power-law dependence, $f(a) \propto a^{-q}$, with the power-law index 
q as a fitting parameter. We attempted the values 2.5, 3.0, 3.25, and 3.5. The minimum and maximum grain sizes are set to
0.01 and 10$^4 \, \mu$m. The smallest grains are needed to have sufficient opacity at the stellar maximum, while the inclusion of
large grains is based on the findings listed above, and the confirmation that for 89\,Her there is also a non-negligible mm flux. 
To implement the grain-size-dependent settling in practice, the 
continuous size distribution is divided into a finite number of bins. The grain radius of each bin is defined as the logarithmic mean of 
the minimum and maximum grain size within the bin. Two bins are used per order of magnitude in grain radius, and in each bin the opacities 
are calculated as the average over 10 sizes to avoid introducing strong resonant features. The turbulent mixing strength is fixed at the standard
value of 0.01 \citep{2012AAMulders}. The same goes for the gas/dust ratio, which is taken to be 100.

One of the largest unknowns in modeling disks is the radial dependence of the surface density distribution, which is typically 
parameterized with a simple power law $\Sigma \propto r^{-p}$. Initially, we do not deviate from the single power-law 
formalism, but do explore a range of values for the index $p$. Unlike for 
protoplanetary disks where typically p is found to be $\sim$1.0 \citep[see][for a general overview]{2011ARAAWilliams}, in the post-AGB 
case this index has not yet been constrained with resolved millimeter observations. Moreover, there is no a priori reason to assume 
that it is the same as in protoplanetary disks, given the difference in formation scenario. 

The inner and outer radius of the disk, as well as the total dust mass are also free parameters of our model. For the inner radius, small 
step sizes of 1\,AU are used because our interferometric data are sensitive to this parameter. A firm 
upper limit on the disk's outer radius comes from the limiting resolution of 400~mas in the PdBI maps of \citet{2007AABujarrabal}. The diffraction-limited
images of \citet{2011MNRASLagadec} impose a thermal IR upper size limit of 300~mas. The HST scattered light images of \citet{2008ApJSiodmiak} 
have a resolution of $\sim$75~mas. The two latter constraints do not imply that there is no material beyond the respective radii, 
it simply means that any material that might be there only emits/scatters with a very low contrast at the given wavelengths. Nevertheless, 
these results do indicate that the disk is likely compact. The effect of the outer disk radius on the SED is rather small, 
and we only vary it because it affects the surface density distribution in a different manner than changing the power-law index.

Finally, the stellar properties are also required for the radiative transfer calculations and to determine the vertical structure of the disk. 
Some of these values were discussed extensively and redefined in Paper I. The stellar mass is fixed to 1~$M_{\odot}$, taken as the 
sum of a 0.6~$M_{\odot}$ post-AGB star and a 0.4~$M_{\odot}$ companion, so that the corresponding inclination is $\sim$13.5$^\circ$ 
(derived from the measured mass function of \citet{1993AAWaters}). To correct for interstellar reddening, we use the reddening law 
of \citet{2004ASPCFitzpatrick} and an $E(B-V)=0.07$ (see Paper I). The final list of input parameters and associated 
values are listed in Table~\ref{table:MCMaxInput}.

\begin{table*}
\caption{The input parameters for the MCMax models and the best-fit values found in Sects.~\ref{subsection:fullscattering} and \ref{subsect:nearIR}.
The $\chi^2$ also list between brackets the relative value with respect to the minimum of the corresponding grid. }             
\label{table:MCMaxInput}      
\centering                          
\begin{tabular}{l l c r r r}        
\hline\hline                 
 & Parameter & Value(s) & Best 1 & Best 2 & Best alt.\\    
\hline
\hline                        
 System & d (kpc) & 1.5 & & & \\   
  & i ($^\circ$) & 13.5 & & & \\
\hline
 Stellar & T$_{\mathrm{eff}}$ (K) & 6550 & & & \\      
  & R$_\star$ (R$21_\odot$) & 71.0 & & & \\
  & L$_\star$ (L$_\odot$) & 8350 & & & \\
  & M$_\star$ (M$_\odot$) & 1.0 & & & \\
\hline
 Dust & M$_{\mathrm{dust}}$ (M$_\odot$) & 10$^{-5}$-10$^{-2}$ & 5\,10$^{-4}$ & 5\,10$^{-4}$ & 1\,10$^{-4}$ \\
  & q & 2.5;3.0;3.25;3.5 & 3.0 & 3.0 & 3.25 \\
  & a$_{\mathrm{min}}$ ($\mu$m) & 0.01 & & & \\
  & a$_{\mathrm{max}}$ ($\mu$m) & 10$^4$ & & & \\
  & gas/dust & 100 & & & \\
  & cont. opacity & Fe/C & Fe & Fe & C \\
\hline
 Structure 1 & R$_{\mathrm{in}}$ (AU) & 3-10 & 4.5 & - & - \\
  & R$_{\mathrm{out}}$ (AU) & 50;100;150;300 & 50 & - & - \\
  & p & 0.5;1.0;1.5;2.0 & 0.5 & - & - \\
\hline
 Structure 2 & R$_{\mathrm{in}}$ (AU) & 3.75;4.00;4.25;4.50;4.75 & - & 3.75 & 3.75 \\
  & R$_{\mathrm{out}}$ (AU) & 50 & - & & \\
  & R$_{\mathrm{mid}}$/R$_{\mathrm{in}}$ & 1.25;1.5;2.0;2.5;3.0;3.5;4.0 & - & 3.0 & 4.0 \\
  & p$_{\mathrm{in}}$ & -2.0;-1.0;-0.5;0.0;1.0 & - & -2.0 & -1.0 \\
  & p$_{\mathrm{out}}$ & 1.5 & - & 1.5 & 1.0 \\
\hline 
 $\chi^2$(/min) & SED$_{\mathrm{OPT}}$ &  & 135(1.7) & 157(1.7) & 161(1.2) \\
  & SED$_{\mathrm{NIR}}$ &  & 14.5(9.3) & 2.0(1.3) & 2.6(2.8)\\
  & SED$_{\mathrm{IR}}$ & & 28.6(7.6) & 3.3(3.2) & 4.4(3.9) \\
  & VIS$_{\mathrm{OPT}}$ &  & 5.1(1.6) & 9.6(2.8) & 12.1(2.0) \\
  & VIS$_{\mathrm{H}}$ &  & 22.5(1.5) & 28.5(2.8) & 22.5(2.5) \\
  & VIS$_{\mathrm{K}}$ &  & 21.0(3.3) & 8.4(2.0) & 8.0(2.3) \\
  & VIS$_{\mathrm{MIDI}}$\tablefootmark{a} & & 2.6(2.0) & 2.8(1.2) & 3.5(1.5) \\
\hline                                   
\end{tabular}
\tablefoot{\tablefoottext{a}{Limited to the 8.0-9.0~$\mu$m continuum wavelength range.}}
\end{table*}

\section{First results} \label{section:firstresults}
In this section, we perform a first confrontation between the radiative transfer models and the observations, from which it will become clear that
our model is too simple to fit the observations. In subsequent sections attempts will be made to resolve the discrepancies.

\subsection{An initial selection} \label{subsection:initialselection}
A large grid of MCMax models was calculated with the input parameters listed in Table~\ref{table:MCMaxInput}. Our fitting strategy 
consisted of several steps. First, for computational efficiency, the full grid was calculated under the assumption of isotropic scattering, 
and without a full ray tracing of the spectrum. In the case of isotropic scattering, the raw Monte Carlo spectra can be used as a 
good proxy for the real spectrum, as long as sufficient photons are used. 
For comparison with the photometric observations, we integrate the model spectra to compute synthetic photometry for each of 
the passbands in our data set, taking into account whether a bolometer- or CCD-type of detector is involved. 

Each model in the grid is then first compared to the IR part of the SED. To this end, we compute a normalized 
$\chi^2$ that includes all photometry at wavelengths larger than 6\,$\mu$m.
Only mid- to- far IR fluxes are included in the $\chi^2$ because at these wavelengths the effect of scattering is small. In this way, a first 
selection can be made of models that generally fit the observed IR-excess. A very rough cutoff criterion of 
$\chi^2 < 100 \times \chi^2_{\mathrm{min}}$ was applied. The advantage of this approach is that the correlations that exist between 
different parameters, and that cannot be resolved by SED-fitting, are withheld. 
From inspection of the retained models, we note:
\begin{itemize}
 \item Most models that fit the $\chi^2$-criterion have a dust size distribution power-law index q of 3.0 or 3.25, but a minority of models with 2.5 and 3.5 cannot just be excluded. 
 \item The q = 3.5 models generally underpredict the (sub-)mm fluxes, tend to give a too strong 10~$\mu$m feature, and have too large near-IR fluxes. 
 \item For q = 2.5, the far-IR slopes are too shallow, and dust masses larger than 10$^{-3}$\,M$_\odot$ are required.
 \item The best-fit models with q = 3.0 or q = 3.25 have a dust mass in the range 10$^{-4}$-10$^{-3}$\,M$_\odot$.
 \item The outer radius of the disk is not constrained, although models with radii of 50 and 100~AU more often lead to good $\chi^2$ values.
 \item The inner radius and the power-law index of the surface density distribution are not constrained individually, as expected.  
\end{itemize}

\citet{2007AABujarrabal} derived a total gas mass of $\sim$10$^{-2}$~M$_\odot$ for the central (disk) component that was unresolved in their CO-maps, 
assuming a distance of 1~kpc. This was recently confirmed from single-dish CO observations by \citet{2013AABujarrabalB}. Since we assume a 
larger distance (see Paper I), and because of the inherent uncertainties in estimating the total gas content from
$^{13}$CO lines, we allow our models to have a maximal total mass of 0.1~M$_\odot$, or equivalently a dust mass of 10$^{-3}$~M$_\odot$. The minimum dust
mass needed to fit the SED is 10$^{-4}$\,M$_\odot$. 

\subsection{A full scattering treatment} \label{subsection:fullscattering}
All models that fit the initial $\chi^2$ criterion, and the imposed mass constraints, are now recomputed with full scattering. A 
ray-traced spectrum is now needed, as in the full angle-dependent scattering case, the Rayleigh-Jeans tail of the Monte Carlo 
spectra becomes unreliable. To compute the visibilities, a disk position angle of 0$^\circ$ 
east of north is assumed. Given the small inclination, this value cannot be determined as it has a very small influence. A good check 
on the amplitude of the potential error, is to consider the effective baseline defined by \citet{2008ApJTannirkulam}:
\begin{equation}
 B_{\mathrm{eff}} = B_{\mathrm{proj}} \sqrt{\cos^2(\theta) + \cos^2(i) \sin^2(\theta)}
\end{equation}
with $B_{\mathrm{proj}}$ the actual projected baseline length and $\theta$ the angle between the projected baseline and the major axis of the disk. With
our assumed inclination, the correction term amounts to at most 2\%.

Different $\chi^2$ are defined as a measure for goodness-of-fit. The SED is split into three wavelength domains, based on their dominant circumstellar
emission mechanism: 1) scattering dominated for wavelengths below 1.5\,$\mu$m, 2) thermal emission dominated beyond 6\,$\mu$m, 
and 3) a near-IR transition regime where both mechanisms are equally important and where our visibilities impose strong spatial constraints. 
A different $\chi^2$ is defined for each of these domains. Similarly, separate $\chi^2$ values are computed per passband (R, H, K, and N) 
with measured visibilities. The performance of the model at optical wavelengths will be evaluated in Sect.~\ref{sect:optical}. 
First, we search for models that fit all IR constraints, i.e., the IR-excess as well as the mid- and near-IR visibilities. 

We select good models based on the following criterion: $\chi^2 < 4 \times \max(1.0,\chi^2_{\mathrm{min}})$ for \textit{all} of the data sets. 
Taking a minimal $\chi^2_{\mathrm{min}}$ of 1.0, allows us to avoid excluding too many models in case a particular data set is over-fitted.
On the other hand, not imposing $\chi^2_{\mathrm{min}}=1.0$ throughout allows us to find models that globally fit all data sets, without 
necessarily fitting each data set to high precision. One must keep in mind that there is some variability in the data 
that cannot be reproduced by our model. In particular, the near-IR visibilities show some scatter that is larger than the errors of some of the 
corresponding data, which could be a time-dependent effect (perhaps due to the binarity), the result of spatial/azimuthal flux variations 
(e.g., a low-amplitude spiral pattern in the disk) or just an underestimation of some of the errors.


None of the current models fits our criterion. A closer inspection of the best-fitting models shows that the near-IR visibilities cannot be fitted 
simultaneously with the SED. Basically, the models that generally fit the SED well have an inner rim radius that is too large 
to fit the first lobe of the H and K visibility curves (which dominate the $\chi^2$). In turn, the very 
few models that fit these first lobes better, and therefore come out with good visibility $\chi^2$'s, overpredict the H band flux while underpredicting
the near- to mid-IR fluxes, and therefore do not fit the intermediate nor the long near-IR baselines either 
(see the blue dotted line in Figs.~\ref{figure:firstbestSED},~\ref{figure:MCMaxHvis}, and~\ref{figure:MCMaxKvis}; and ``Best1'' in Table~\ref{table:MCMaxInput}).
Models with small inner radii ($\sim$3.5-4.5~AU) that do predict enough flux beyond $\sim$K band were already removed in the previous section, 
as these usually had extreme near- to mid-IR excesses when computed with isotropic scattering. As such, the remaining models with small inner radii 
generally are those with flat surface density distributions. As a check of our strategy, we recompute 
the 20 best models with an inner radius of 4.5~AU that were previously removed with a full scattering treatment, 
and find that these models do not fit our criterion.

\section{A good fit to the IR data} \label{sect:IR}
To improve on the structural deficiency of our model, we change our surface density distribution parameterization and adopt a double power-law 
formalism (see Sect.~\ref{sect:discussion} for a further motivation and discussion of this choice). Since this new parameterization introduces two 
more fitting parameters, the turnover radius R$_{\mathrm{mid}}$ and the inner density power-law index p$_{\mathrm{in}}$,
we reduce the parameter space for four other parameters: the outer radius and surface density power-law index (p$_{\mathrm{out}}$), 
the total dust mass, and the grain size power-law index q. Given the indications for a very compact disk with large 
grains (see Sect.~\ref{subsection:setup}), but only a moderate mass (see Sect.~\ref{subsection:initialselection}), and since 
a significant scale height is required to explain our optical
observations, we adopt our most compact configuration of R$_{\mathrm{out}}$\,=\,50\,AU. Furthermore, we adopt the ``canonical''
value of p$_{\mathrm{out}}$=1.0 or 1.5. The dust mass is limited to 1.0, 2.5 or 5.0~$\times 10^{-4}$~M$_\odot$. The value of q can be 
either 3.0 or 3.25. Based on our geometric analysis of Paper I, and the good correspondence with the first-lobe visibilities of models with inner 
radii of $\sim$4--5~AU in the previous section (see Figs.~\ref{figure:MCMaxHvis} and~\ref{figure:MCMaxKvis}), we vary the inner radius 
between 3.75 and 4.75\,AU, in steps of 0.25\,AU. 

\begin{figure}
   \centering
   \includegraphics[width=8cm,height=7cm]{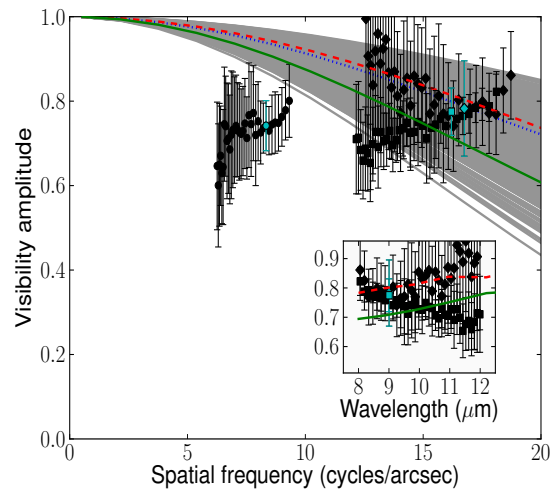}  
     \caption{The three baselines of MIDI data overplotted (at a single continuum wavelength, indicated in cyan) with all double power-law MCMax 
     models computed in Sect.~\ref{sect:IR}. 
     The blue dotted line, ``Best 1'' in Table~\ref{table:MCMaxInput}, is the best single power-law MCMax model in terms of 
     fitting the near-IR visibilities (see Figs.~\ref{figure:MCMaxHvis} and~\ref{figure:MCMaxKvis}). The red dashed line 
     corresponds to the best-fitting double power-law model with the metallic Fe opacity source 
     (``Best 2'' in Table~\ref{table:MCMaxInput}). The green full line shows the best-fitting double power-law model 
     when metallic Fe is replaced by amorphous C (``Best alt.'' in Table~\ref{table:MCMaxInput}). See Sects.~\ref{subsection:fullscattering},
     ~\ref{subsect:nearIR}, and~\ref{sect:carbon} for a discussion of the models. 
     The inset shows the data at the 30~m baselines as a function of wavelength, with the same 
     double power-law models overplotted as in the main figure.
     }
     \label{figure:MIDIvis}
\end{figure} 

\begin{figure}
   \centering
   \includegraphics[width=8cm,height=7cm]{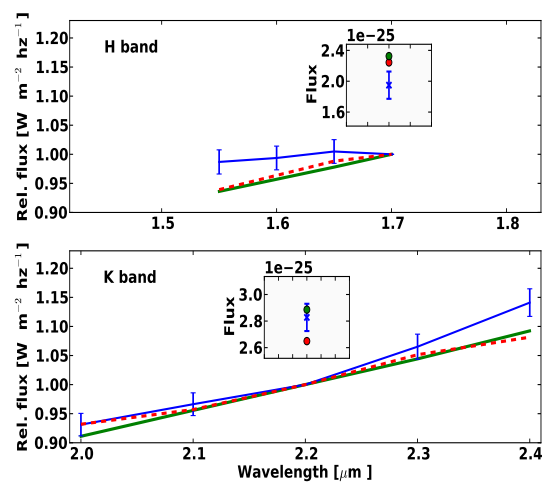}  
   \caption{The PTI absolute fluxes (in the insets) and spectral shapes over the H (upper panel) and K (lower panel) bands. The data are 
   shown in blue. The red dashed and green full lines correspond to the same models as in Fig~\ref{figure:MIDIvis}.
   }
   \label{figure:PTIfluxes}
\end{figure} 

\begin{figure*}
   \centering
   \includegraphics[width=16cm,height=12cm]{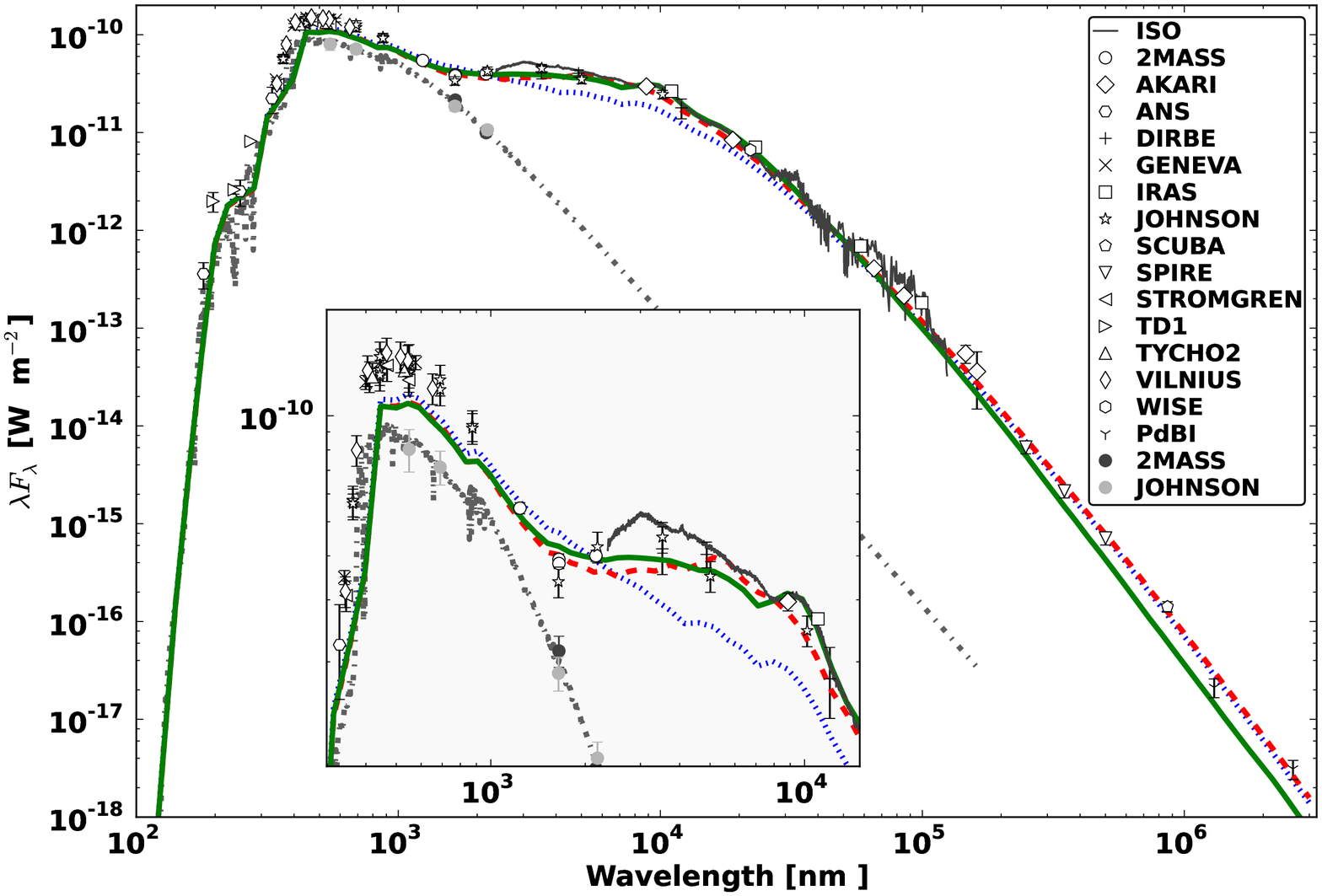}
   \caption{The SED of 89 Her with the same three models as in Figs.~\ref{figure:MIDIvis} and~\ref{figure:PTIfluxes} overplotted. 
     The gray dot-dashed line shows the input Kurucz model, 
     here reddened with an interstellar contribution of $E(B-V)=0.07$ (see Paper I). The black and gray data points are the ``stellar'' 
     V, R, H, and K band fluxes as determined in Paper I. Open symbols represent actual measured fluxes in the passbands that are listed 
     in the legend. The full black line is the ISO spectrum. The blue dotted line, ``Best 1'' in Table~\ref{table:MCMaxInput}, clearly 
     does not agree with the observed SED. A better, but still not excellent, fit to the SED can be obtained with an inner radius twice as 
     large, in combination with $p=1.5$, but this results in first-lobe near-IR visibilities that are too small (see the text). The red dashed and green full
     line correspond to the best-fitting double power-law models with metallic Fe and amorphous C opacity sources, respectively 
     (see Fig.~\ref{figure:MIDIvis}). The MCMax models have been reddened similarly as the Kurucz model.
     }
     \label{figure:firstbestSED}
\end{figure*} 

\subsection{The SED and near-IR visibilities} \label{subsect:nearIR}
The good models are searched in the same way as described in the previous section. Five models in the grid now conform to our acceptance criterion, 
and in general the $\chi^2$'s have decreased significantly below the values of the single power-law models.
Interestingly, all good models have $q=3.0$ (imposed by the H band visibilities) and a dust mass equal to $5 \times 10^{-4}$~M$_\odot$. 
There are no clear preferences for the surface density gradients. The inner and turnover radius 
are also not well defined, but this can be expected since we already refined the sampling in these parameters. In any case,  
the turnover radius lies about three times further than the inner radius of the disk on average. 

In the 2-6~$\mu$m range, a flux deficit is present when compared to the ISO 
spectrum, however, based on the photometry, the fit and the corresponding $\chi^2$ is very good. The H and K band fluxes fit quite 
well, as is confirmed by the PTI fluxes in Fig.~\ref{figure:PTIfluxes}. The slope of the spectrum does not exactly agree: 
there is slightly too much flux in the long-wavelength end of the H band and too little flux in the entire K band. Also, the spectral 
slope towards 2.4~$\mu$m is too shallow with respect to the data, which confirms the mismatch with respect to the short-wavelength 
end of the ISO spectrum. We come back to these small discrepancies in Sects.~\ref{sect:carbon} and~\ref{sect:discussion}.

\begin{figure*}
\centering
   \includegraphics[width=8cm,height=7cm]{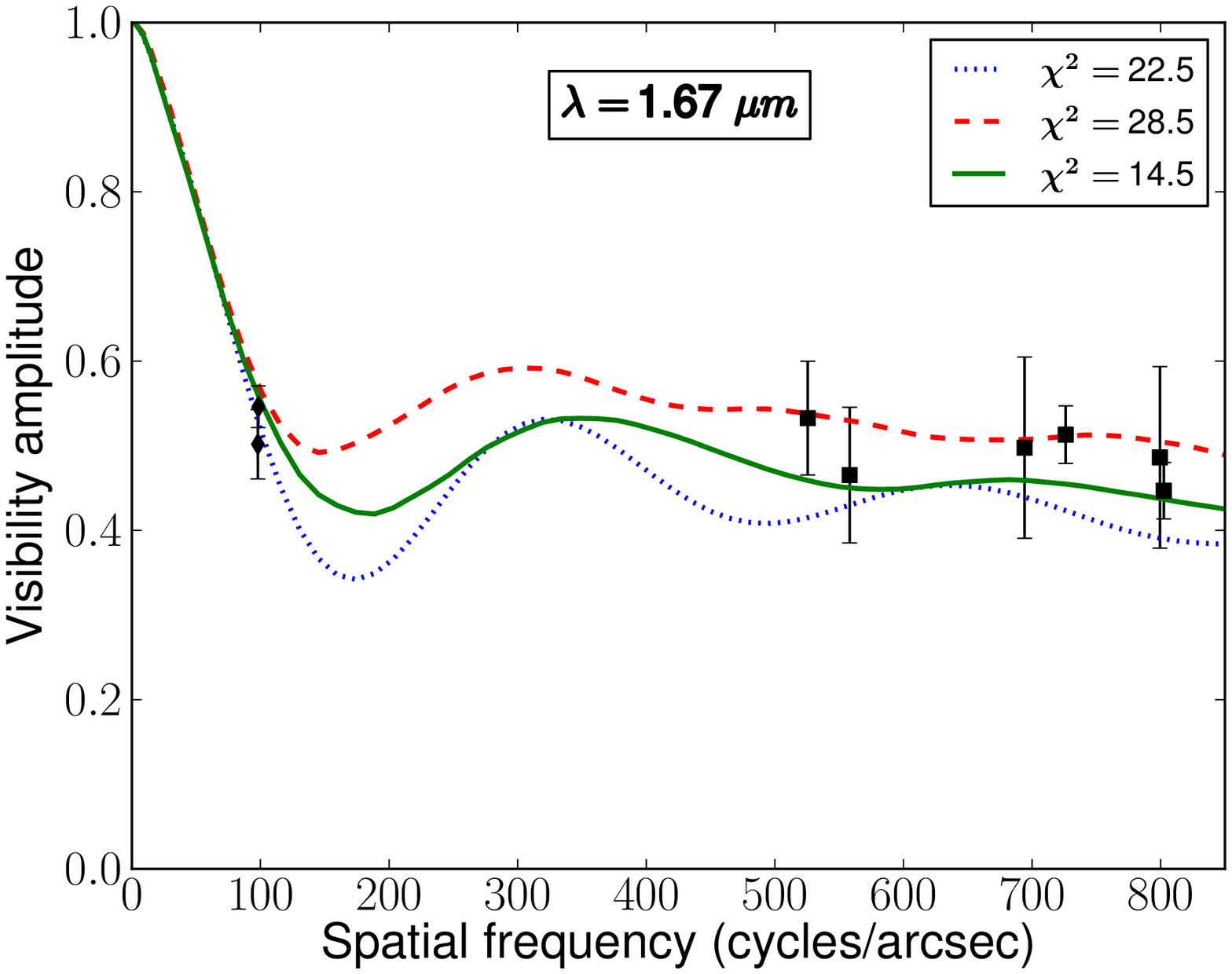} 
   \hspace{1cm}
   \includegraphics[width=8cm,height=7cm]{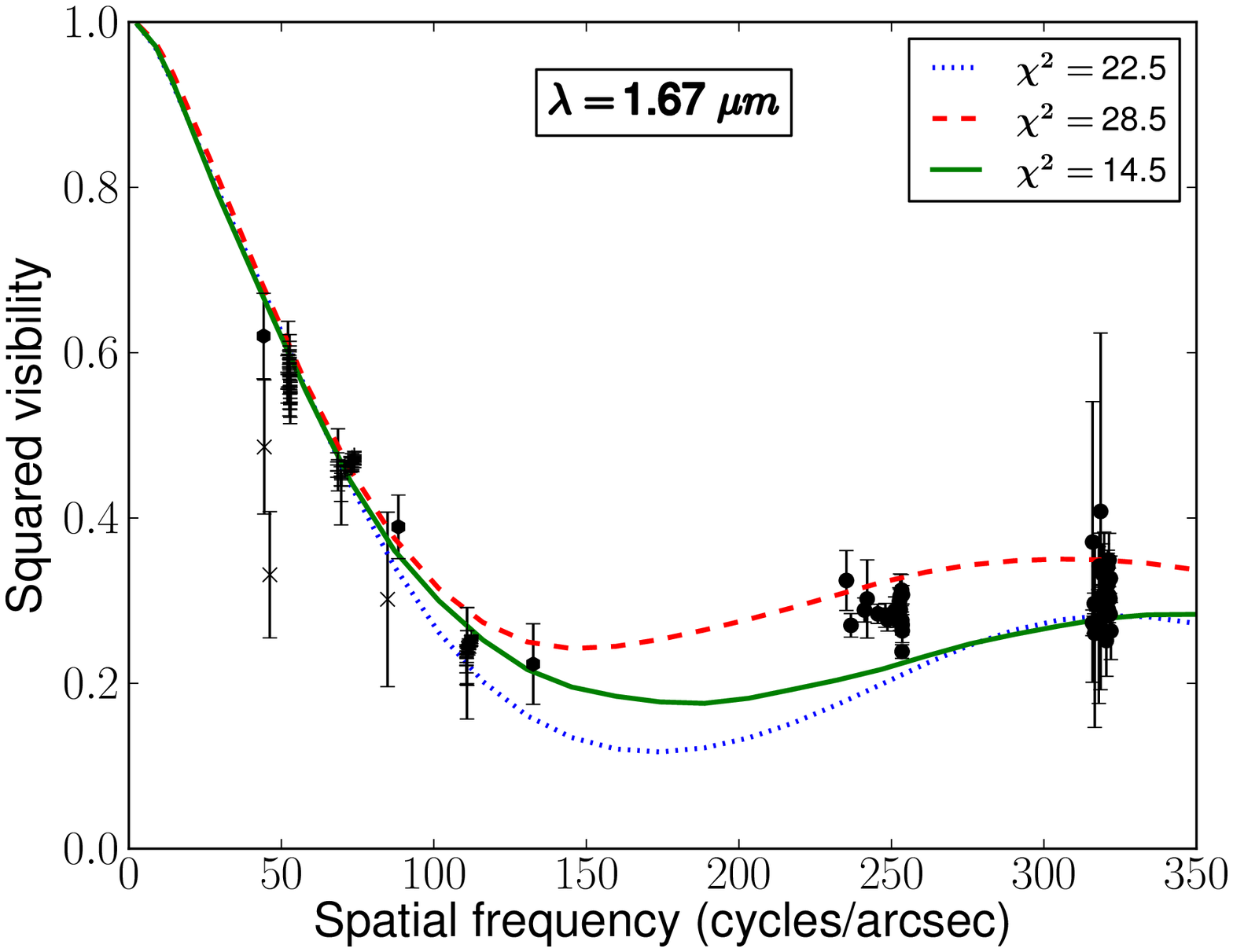} 
     \caption{H band interferometric data. The left and right panels show the H band visibility amplitudes and squared visibilities
     observed by the various instruments. The visibility amplitudes, as produced by the data reduction 
     software packages of the CLIMB and CLASSIC instruments, are not squared in order to preserve the original error statistics.
     The data are the same as in Paper I. The same three models are shown as in Figs.~\ref{figure:MIDIvis} and~\ref{figure:firstbestSED} 
     and as listed in Table~\ref{table:MCMaxInput}.}
     \label{figure:MCMaxHvis}
\end{figure*}

\begin{figure*}
\centering
   \includegraphics[width=8cm,height=7cm]{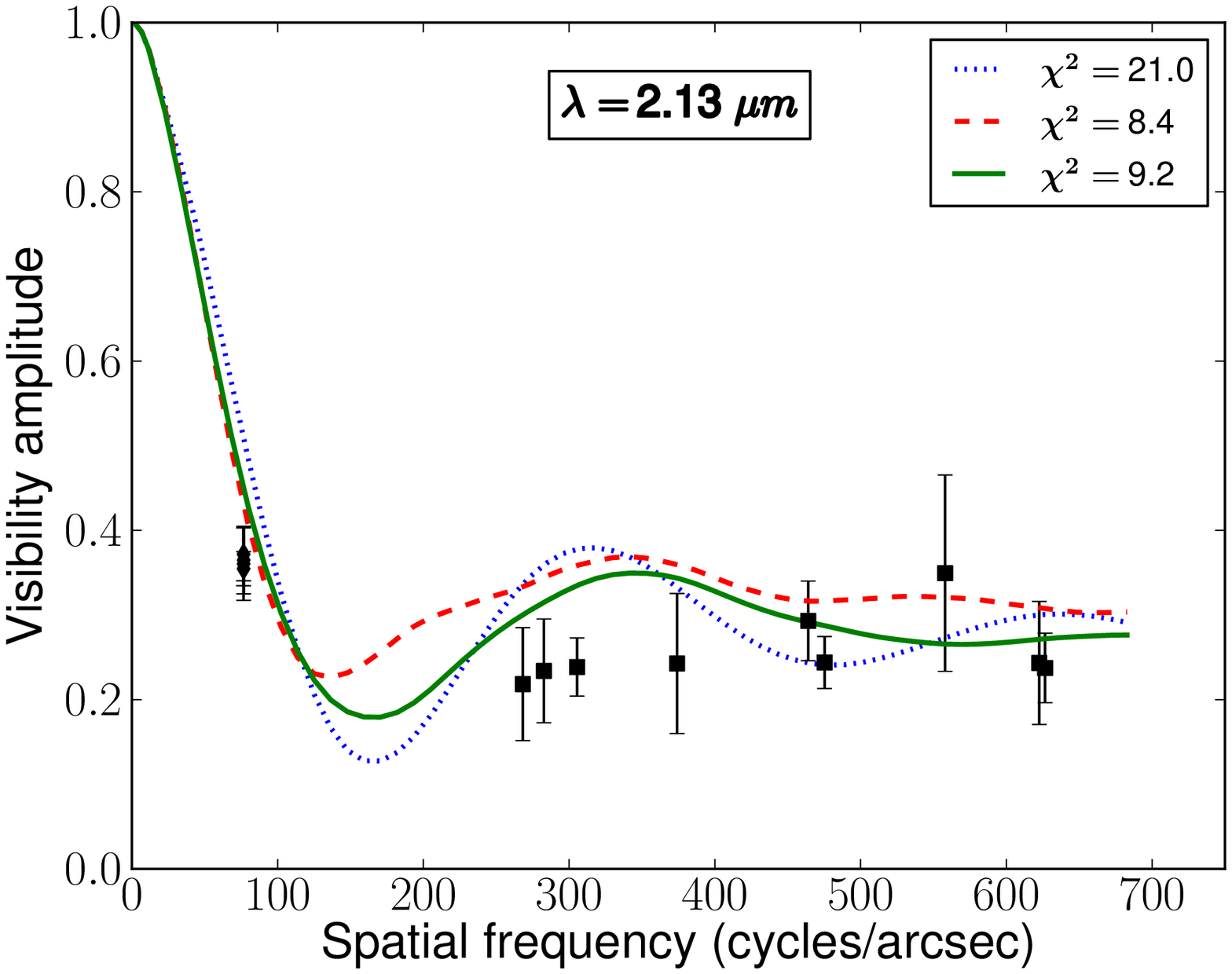} 
   \hspace{1cm}
   \includegraphics[width=8cm,height=7cm]{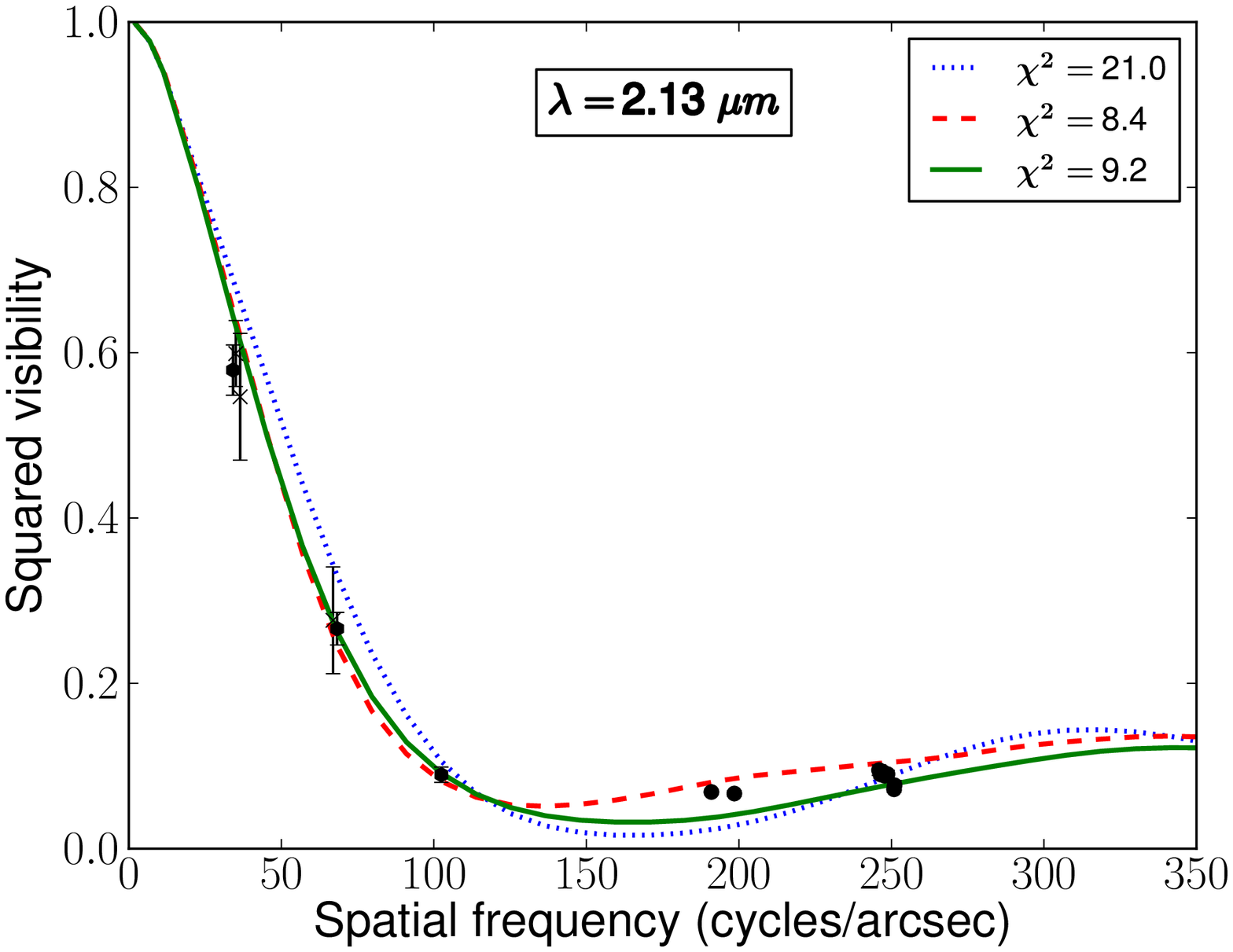} 
     \caption{The same as Fig.~\ref{figure:MCMaxHvis} but for the K band.}
     \label{figure:MCMaxKvis}
\end{figure*}

The model fits the visibilities (shown in Figs.~\ref{figure:MCMaxHvis} and~\ref{figure:MCMaxKvis}) quite well in both the H and K band. 
The rather high $\chi^2$ in the H band suggests otherwise, but this can largely be attributed to the previously mentioned scatter in the data. 
The largest contribution to the H band residuals comes from the IOTA data around spatial frequencies of 115 cycles/arcsec 
and the PTI data near 325 cycles/arcsec. 

The parameter values of the preferred model are listed in Table~\ref{table:MCMaxInput} as ``Best 2,'' 
and we accept it as the model that fits our IR data sets best. The structure of this model will be examined in more detail in Sect.~\ref{sect:discussion}.

\begin{figure*}
   \centering
   \includegraphics[width=8cm,height=7cm]{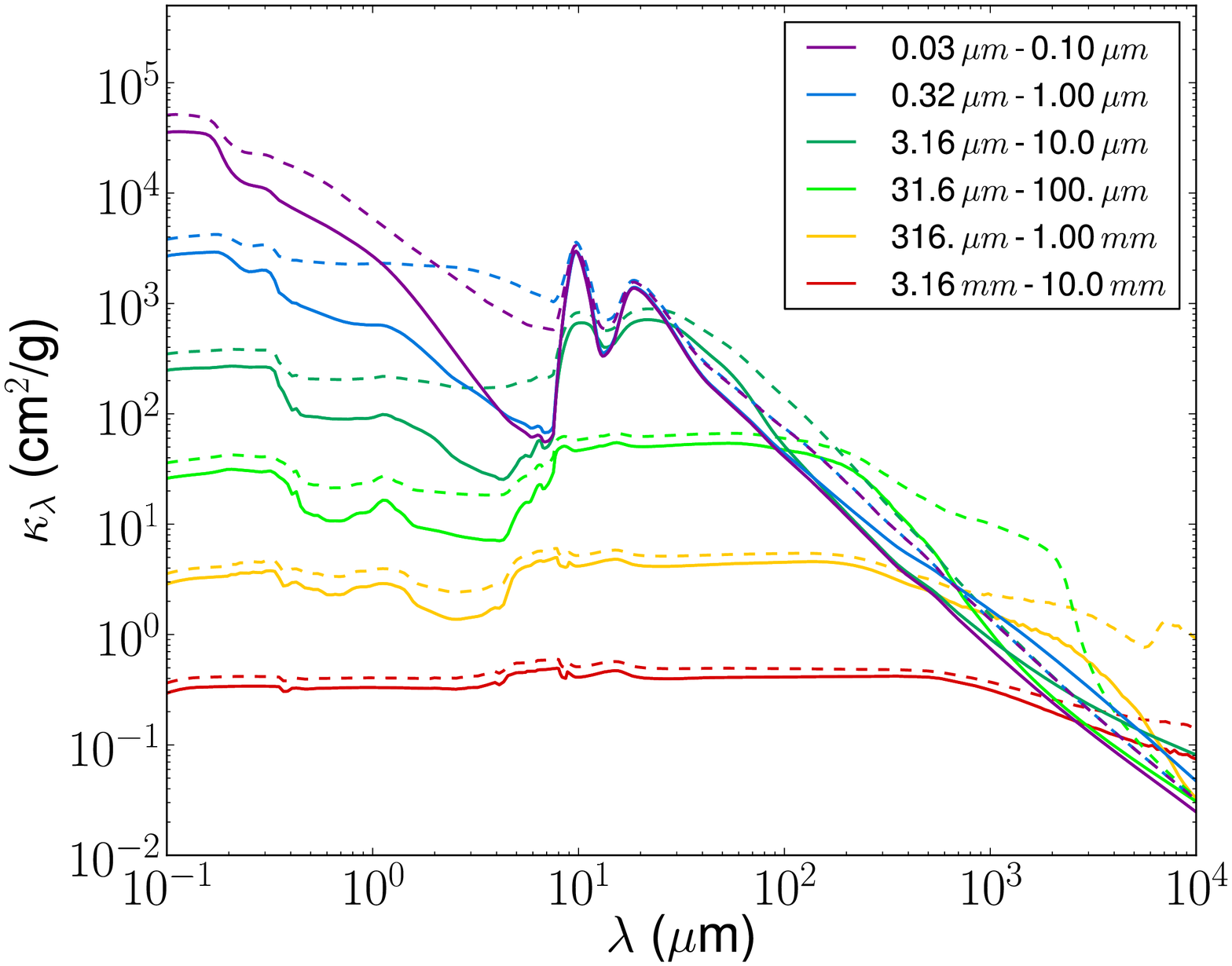}  
   \hspace{1cm}
   \includegraphics[width=8cm,height=7cm]{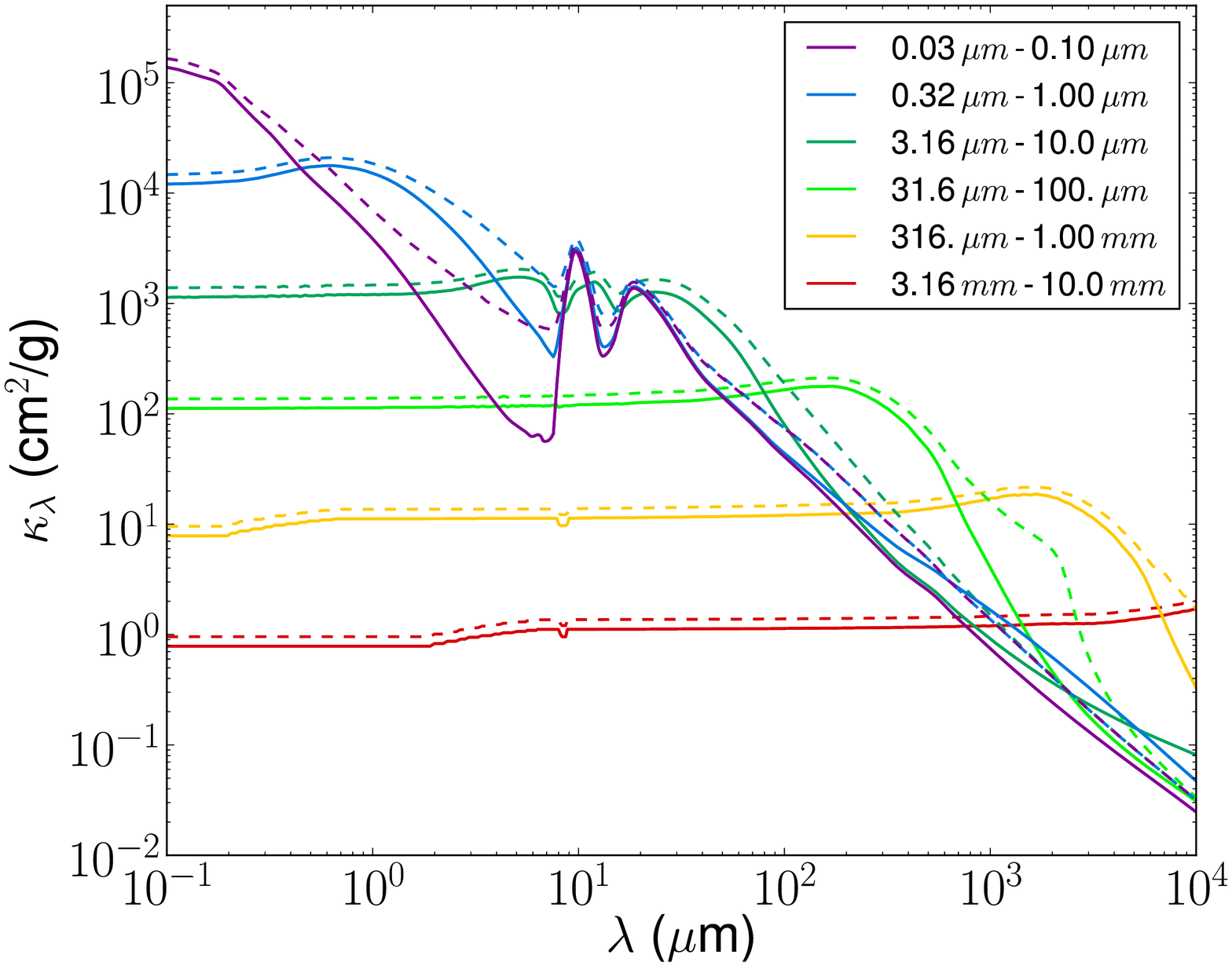}
   \caption{The mass absorption (left) and total extinction (right) coefficients of the mixtures used in this work. Our 'standard' mixture (the full lines) 
   contains 15\% metallic iron, while the alternative mixture (the dashed lines) has 15\% amorphous carbon instead. The latter composition is exactly the 
   same as used by \citet{2012AAMulders}. Only six grain sizes are shown for clarity.}
   \label{figure:opacities}
\end{figure*} 

\subsection{The mid-IR visibilities}\label{sect:midIRMCMax}
The MIDI data have an insufficient spatial resolution to strongly constrain the position and shape of the inner disk rim, and the data are 
more sensitive to the bulk of the inner disk. Therefore, it is surprising that the visibilities (see Fig.~\ref{figure:MIDIvis}) are 
so similar at the 15, and 30\,m baselines. The predictions of all our double power-law models and of the ``best''
single power-law model are depicted as well. Clearly, none of the models simultaneously fits  
both baselines, or even just the slope between them. Nevertheless, basically all of the models fulfill our acceptance criterion 
and have a good $\chi^2$, thanks to the larger weight of the 30\,m baseline (as it is observed twice) and the rather large errors of the data.
The shape of the MIDI visibility spectrum does not hold much information, given that this shape varies between the two 30\,m observations.
While our best-fit double power-law MCMax model is not perfect, it formally reproduces the mid-IR data sufficiently. 
In Sect.~\ref{subsection:midiDiscussion}, we return to the 15\,m discrepancy.

\section{Amorphous carbon?} \label{sect:carbon}
Finally, we also computed a grid of models, with the same parameter values as in the previous section, but with 15\% of amorphous carbon 
instead of metallic Fe as the additional opacity source (see Sect.~\ref{subsection:compositiondiscussion} for a discussion of our 
motivations). As is illustrated in the left panel of Fig.~\ref{figure:opacities}, 
amorphous C has a higher mass absorption coefficient in the near- to mid-IR domain compared to metallic Fe, independent of the grain size, and 
in the (sub-)mm domain for the large grains. In terms of total extinction
(as shown in the right panel), however, the differences are much smaller, except for the smallest grain sizes. Therefore, one might 
expect a better agreement of these models to the ISO spectrum and the exact shape of the SED in the near-IR. 

We applied the same strategy to find the best-fit models within this grid (see also Table~\ref{table:MCMaxInput} and Figs.~\ref{figure:MIDIvis},
~\ref{figure:PTIfluxes},~\ref{figure:firstbestSED},~\ref{figure:MCMaxHvis}, and~\ref{figure:MCMaxKvis}). The near-IR 
fluxes are reproduced marginally better, as are the mid-IR fluxes, the silicate feature at 10~$\mu$m, and the H band visibilities. 
On the other hand, the fit to the (sub-)mm fluxes degrades. The reason for these changes lies in the changed values of several parameters: 
the models are now allowed to have relatively more small grains ($q=3.25$), but only if the 
total dust mass (M$_{\mathrm{dust}}=1\,10^{-4}$) decreases accordingly. 
The surface density is now also generally flatter with p$_{\mathrm{out}}=1.0$.

\section{Fit to the optical data} \label{sect:optical}
\begin{figure}
   \centering
   \includegraphics[width=8cm,height=7cm]{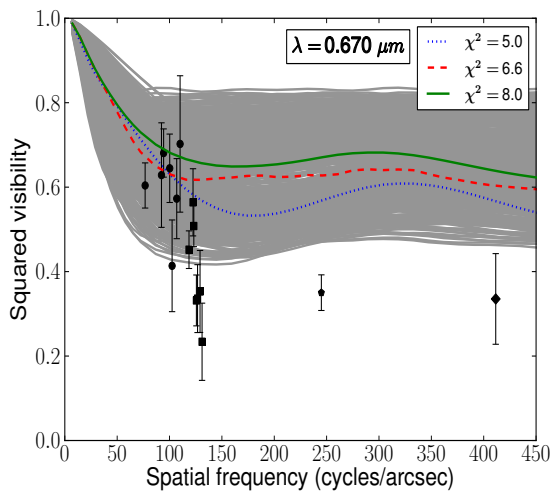}  
   \caption{The NPOI and VEGA squared visibilities at 673\,nm. Our complete grid of double-power-law MCMax 
   models is also displayed (see Sect.~\ref{sect:IR}). The best-fit models listed in Table~\ref{table:MCMaxInput}
   are shown with the same color coding as in Figs.~\ref{figure:MIDIvis}, \ref{figure:firstbestSED}, \ref{figure:MCMaxHvis} and~\ref{figure:MCMaxKvis}.}
   \label{figure:opticalVis}
\end{figure}

We finally confront the models with the optical visibilities, and the optical part of the SED. Fig.~\ref{figure:firstbestSED} 
already illustrated that the best IR models of the previous sections fail at wavelengths below 1~$\mu$m. This is 
the case for \textit{all} of the calculated models. Of all the models computed in Sect.~\ref{sect:IR}, the model that performs best
still has a $\chi^2 \sim 80$ for the optical part of the SED, and underpredicts the observed scattered flux. To illustrate 
this more clearly, Fig.~\ref{figure:opticalVis} shows the NPOI and VEGA visibilities at 673\,nm, with all of the full-scattering models that were 
calculated in Sect.~\ref{sect:IR} overplotted. 

\begin{figure}
 \centering
 \includegraphics[width=8cm,height=7cm]{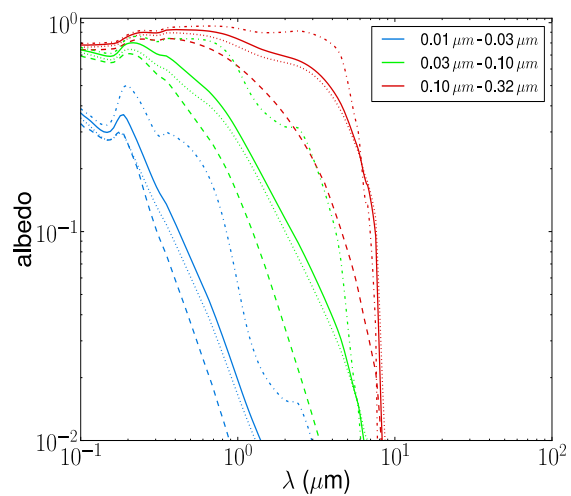}  
 \caption{The albedo of the four smallest particle sizes and for four different compositions: silicates + amorphous carbon (dashed lines), 
 only silicates (dotted lines), silicates + 15 \% iron (full lines), and silicates + 30 \% iron (dot-dashed lines).}
 \label{figure:albedos}
\end{figure}

Not a single model reaches the low visibilities that were observed. Interestingly, the best model of Sect.~\ref{subsection:fullscattering} (the 
blue dotted curve on Figs.~\ref{figure:firstbestSED} and~\ref{figure:opticalVis}) performs better than our best double power-law models computed in the 
previous sections. Models with amorphous C generally perform worse at optical wavelengths than models with metallic Fe, due to their smaller 
single-scattering albedo (see Fig.~\ref{figure:albedos}) and larger fraction of small grains.
For our metallic Fe models, much higher fractions of optical scattered flux are found on average with single rather than with double surface density power laws. 
These models have large dust masses, however, often in combination with a large fraction of small grains ($q=3.25$), so they are 
totally incompatible with the IR constraints as they predict too much H, K and often even mid-IR, fluxes. The best double power-law models, with a 
$VIS_{opt}$ $\chi^2$ of only 3.1, suffer in fact from the same problem, which is most pronounced in the H band. A closer inspection
shows that the lowest $\chi^2$ actually does not correspond with the highest amount of optical scattered light, which is reserved for the models with 
large turnover radii R$_{mid}$. The latter however underpredict the visibilities at the short spatial frequencies, which have a larger weight in 
the $\chi^2$ simply by their number, in particular in the channels near 0.85~$\mu$m (not shown in Fig.~\ref{figure:opticalVis}). Basically, 
these models have a larger area over which the disk surface contributes to the scattered light (see also the $\tau_V = 1$ surface in 
Figs.~\ref{figure:bestModelTStructure} and~\ref{figure:bestModelDensityStructure}). Since the short spatial frequencies are fairly 
well reproduced by the majority of the models, the resulting range in $\chi^2$ is rather modest: the worst model only has a $\chi^2 \sim 22$.
Finally, we note that many models, including our general best-fit model, result in very flat visibility plateaus at the high spatial frequencies, as
is suggested by the data as well.

\section{Discussion} \label{sect:discussion}
In this work, we have shown that with reasonable input parameters, and using a self-consistent treatment of dust settling, radiative transfer disk models 
can be constructed that are in good agreement with all our infrared observations within their uncertainties. The models 
successfully reproduce the general trend in the near-IR visibilities and fit the full SED up to sub-mm wavelengths very well. 
On the other hand, these best models, and in fact every one of our models, are incapable of producing a sufficient amount of optical scattered light. 
Before we discuss the deficiencies, we put our results in perspective.

\subsection{The inner rim curvature}
That a simultaneous fit to the SED and the near-IR visibilities could not be found with a single power-law model is not surprising.
Our near-IR interferometric observations clearly prefer an extended and smooth emission region (see also Paper I), not the wall-like 
emission profile resulting from a single power-law parameterization of the surface density. For protoplanetary disks, interferometric 
investigations have already extensively shown that the inner disk rim is usually not sharp, but has a smooth geometry 
\citep[see][for a nice review]{2010ARAADullemond}. Different physical processes have been identified to cause this rounding off. 

\citet{2005AAIsella} showed the importance of including an accurate physical description of dust sublimation and evaporation, and 
in particular its temperature dependence on the gas pressure. \citet{2007ApJTannirkulam} for the first time presented a 2D radiative transfer 
model of a rounded-off rim by considering the effect of dust growth and settling. Due to these effects, big grains, which survive closer to the star thanks
to their ability to cool more efficiently, are also primarily found in the midplane, and thereby round off the rim.

The results of \citet{2005AAIsella} were later confirmed by a more extensive exploration of the effect of global disk properties on inner rim models, 
using a detailed physical treatment of dust formation and destruction in 2D radiative transfer models \citep[][]{2009AAKama}. An important 
assumption in the latter work is their double power-law radial parameterization of the surface density, as assumed here. Their smooth 
inner structures, often having wide optically thin dust regions in front of the actual rim, 
can partly be attributed to this parameterization, which is probably more realistic than the wall resulting from a single power-law model, as we started 
with in Sect.~\ref{section:firstresults}.

Our models are not computed with a sophisticated dust sublimation/evaporation treatment, like in \citet{2009AAKama}, since 
it requires an immense computation time to do so for a grid of models. The self-consistent inclusion of a grain size distribution, 
combined with dust settling, already increased the computation time of a single model from tens of minutes to several hours (on a single CPU).
\citet{2009AAKama} found smaller inner rim radii than \citet{2005AAIsella} and \citet{2007ApJTannirkulam}. This was attributed to
the implementation of an abundance correction to the used dust sublimation law by \citet{2009AAKama}.
It would be interesting to explore how these detailed inner rim curvature calculations exactly depend on the 
abundances, and what would be the effect in a post-AGB context. In particular, it would be interesting to investigate how our 
best-fit models would evolve if dust destruction were properly included in the calculations, but this is beyond the scope 
of the current paper. We also note that \citet{2009AAKama} did not use a grain size distribution for their
dust, and hence neglected the effect of settling.

Instead of extending the model physics, we preferred in this paper to do a wider exploration of the model parameter space, 
since post-AGB disks have not yet been studied extensively. In the following subsection we check,
a posteriori, the temperature and density structure of our best-fit models.

\subsection{The disk structure}
\begin{figure*}
\centering
   \includegraphics[width=17cm]{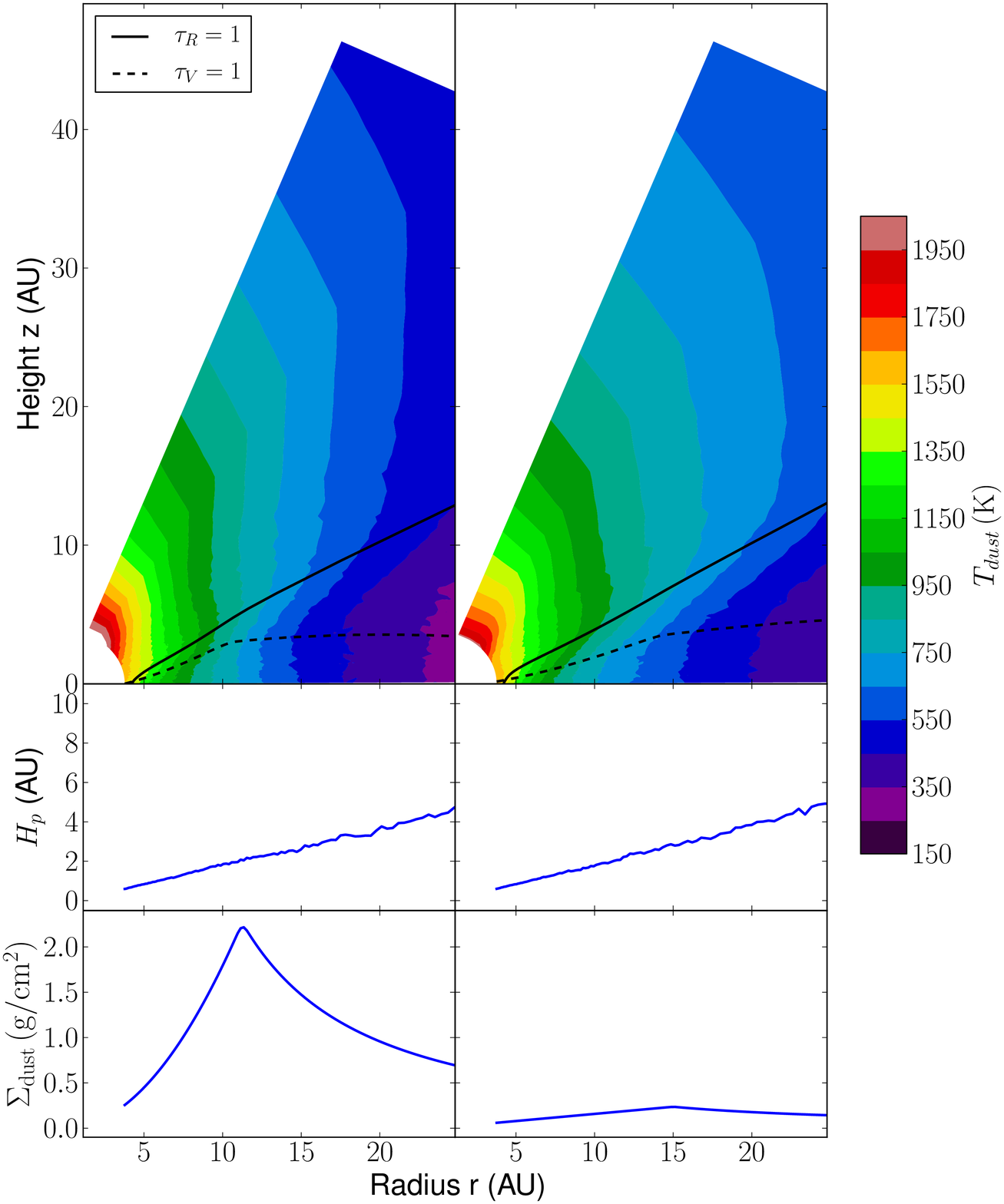}  
   \caption{Temperature structure of our best-fit models. The upper panels show maps of the temperature, with the radial and 
   vertical $\tau = 1$ surfaces at 0.55~$\mu$m overlaid as solid and dashed lines, respectively. The middle and lower 
   panels display the midplane pressure scale height and integrated dust density, respectively, as a function of the disk's radius. 
   On the left and right, the best Fe and amorphous C containing models are shown, respectively.}
   \label{figure:bestModelTStructure}
\end{figure*}

Fig.~\ref{figure:bestModelTStructure} shows the temperature structure of our best-fit Fe (left) and amorphous C (right) containing models 
as a function of the radial and vertical position in the disk, overplotted with the radial and vertical $\tau = 1$ surfaces 
at 0.55~$\mu$m. The figure also shows the radial profile of the gas pressure scale height in the midplane (middle panels) and the 
radial profile of the vertically integrated surface density (lower panels). 

\begin{figure*}
\centering
   \includegraphics[width=17cm]{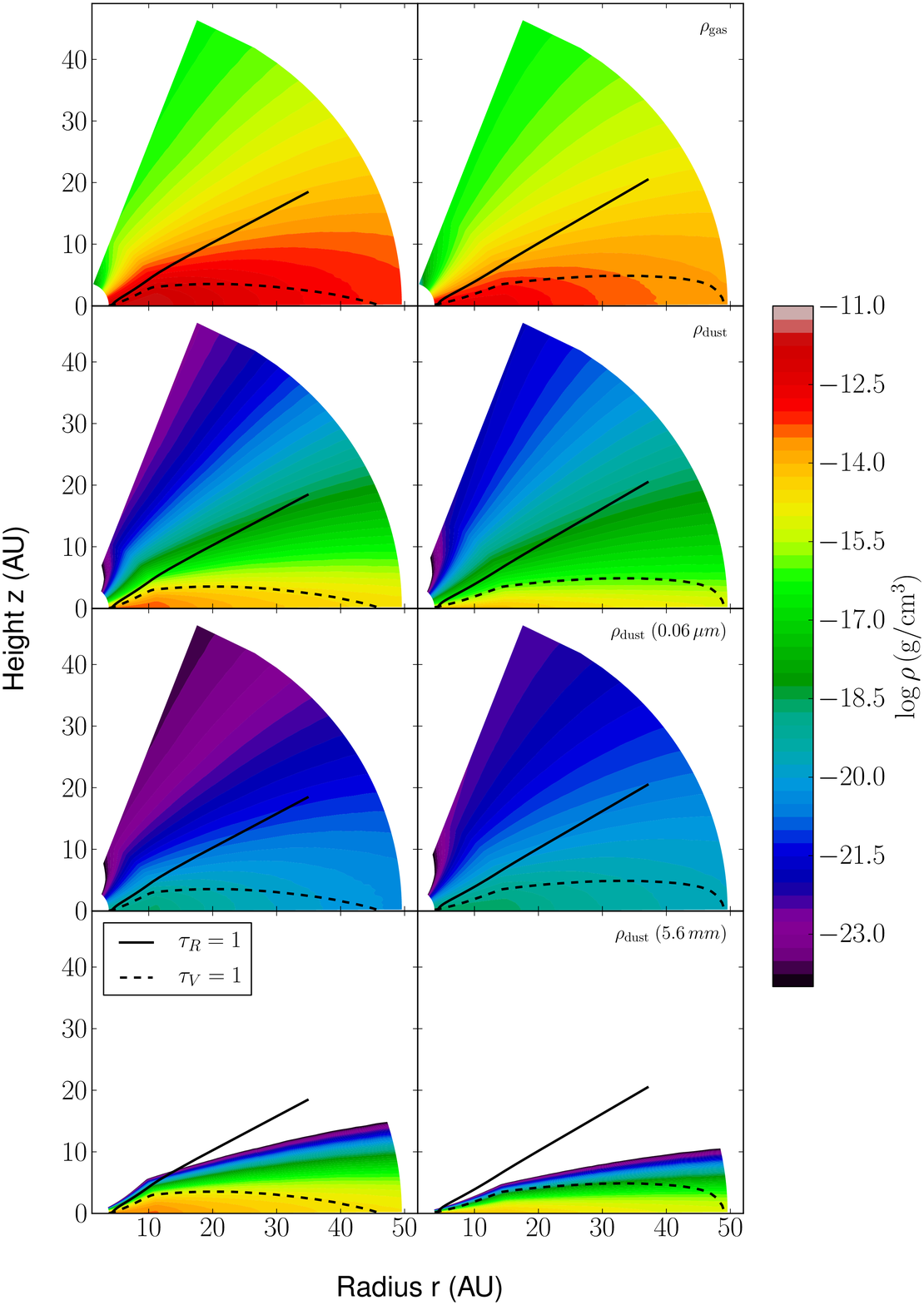}  
   \caption{Density structure of our best-fit models. From top to bottom, maps of the gas density, the dust density, the dust density for particles of 
   $\sim$0.06~$\mu$m, and the dust density for particles of $\sim$5~mm are shown. On all panels the radial and 
   vertical $\tau = 1$ surfaces at 0.55~$\mu$m are overlaid as well, with solid and dashed lines, respectively. 
   The left and right panels contain the best Fe and amorphous C containing models, respectively.}
   \label{figure:bestModelDensityStructure}
\end{figure*}

An inspection of the inner rim dust temperatures points out that they are rather high, up to $\sim$1600~K. The model with amorphous C has a slightly 
smaller temperature at the inner rim, but only by $\sim$50~K. Interestingly, the model with metallic Fe also has a larger radial temperature gradient 
compared to the amorphous C model. High dust temperatures have appeared in the literature before. \citet{2010AABenisty} even needed 
values up to 2200~K, but these values may not be realistic. Considering the results of \citet{2009AAKama}, who attempted to implement the most
accurate treatment of dust sublimation possible, and adopting our gas density of $\leq$10$^{-11.5}$~g/cm$^3$ 
(see Fig.~\ref{figure:bestModelDensityStructure}), sublimation temperatures of only 1300~K for olivine and 1400~K for iron are to be expected. 
It can therefore be foreseen that by including dust sublimation properly, our inner rim radius would increase, which would be 
incompatible with the near-IR interferometric data.

The gas as well as dust density structure of the same two models are depicted in Fig.~\ref{figure:bestModelDensityStructure}. The figure also separately
shows the density of small (i.e., 0.06~$\mu$m) as well as large (i.e., 5.6~$\mu$m) grains, to illustrate the relevance of dust settling.
The Fe containing model in general has higher densities than the C models, except in the very small grains. This is expected, 
given the differences in total dust mass and grain size distribution (see Table~\ref{table:MCMaxInput}).

We fixed some parameters and made some implicit assumptions in our modeling. These specific choices could impact our results.
Our total dust mass agrees with the value determined by \citet{2013AABujarrabalB} based on single-dish, rotationally-excited 
CO line fluxes, in particular for our C-containing model. If the true disk outer radius would be larger than the 50~AU assumed in our 
double power-law models, however, the required dust mass may increase as well. On the other hand, the mass derived by \citet{2013AABujarrabalB} needs to be 
validated with full radiative transfer calculations of the line profiles, taking the disk structure into account. 
Another parameter could be the turbulent mixing strength $\alpha_{\mathrm{turb}}$, which determines the efficiency of vertical mixing, 
and hence strongly impacts on the vertical structure of the disk. Similarly, the central star's mass and the global dust/gas ratio influence 
the disk's scale height. We calculated a few models with $\alpha_{\mathrm{turb}}$ increased or decreased by an order of magnitude
or with M$_\star$\,=\,0.75\,M$_{\odot}$, and it is unlikely that these parameters will affect our main results.

\subsection{The MIDI 15\,m baseline} \label{subsection:midiDiscussion}
The discrepancy between our models and the 15\,m MIDI baseline could be because of the poorer atmospheric 
conditions during the night of that measurement. If this is not the case, however, then the smoothness of \textit{all} of 
our model visibility curves shows that to reproduce both observations, a simple continously declining radial 
density profile does not work. Instead, a strong decrease in the optical depth of the disk just behind the inner rim 
is needed. This means that to fit these data, a gap must be developing in the disk between $\sim$10 and 
40~mas in radius, or that there must be a strong shadow that is not present in our MCMax models.

There is an ongoing debate about shadows and puffed-up inner rims in protoplanetary disks \citep[see, e.g.,\,the review of][]{2010ARAADullemond}.
In some of the original models \citep[e.g.,\,those of][]{2004AADullemond}, the region immediately behind the inner rim had a smaller 
scale height than expected from a fully flaring disk. The interpretation was that this was because of a strong puffing-up of the 
frontally illuminated inner rim, which then casts a shadow on the region immediately behind it. It was, however, questioned whether 
such a shadow could lead to a (partial) collapse of the outer disk \citep[see, e.g., the discussion in][]{2008NewARWood}. 
Instead, it was suggested that the very steep surface density gradient in the self-shadowed models is responsible for 
the disk not reaching the large scale heights of flaring disks. Also, the too simple treatment of scattering and radial 
radiative diffusion of heat might have played a role in the early models, which is no longer a problem in the more recent 
Monte Carlo based radiative transfer codes.

Based on the \citet{2004AADullemond} models, \citet{2005AAvanBoekel} presented simulated visibility curves at 10~$\mu$m 
of a flaring (with partial ``shadowing'') and fully self-shadowed case. Interestingly, these curves have a plateau in 
the continuum at 7.7~$\mu$m because of the presence of a shadow in the models. Despite that our observations look 
very similar in the continuum, the inadequate treatment of scattering and diffusion in these models cannot 
be neglected. Therefore, if real, a gap interpretation to explain our observations would seem more likely. 

The SED does not show any sign of a gap in the disk, while for protoplanetary (pre-)transition disks, in which gaps have been 
spatially resolved, this is mostly clearly the case \citep{2011ARAAWilliams}. For this reason, we assume that our 15\,m observation 
is untrustworthy, but more mid-IR interferometric observations are needed to confirm this.

\subsection{The dust composition}\label{subsection:compositiondiscussion}
Our choice to explore models with amorphous C was inspired by the recent result of \citet{2013AAAcke}, and by the need 
to investigate whether a change in dust composition can significantly affect the predicted amount of scattered light 
by the disk (see Sect.~\ref{section:scatteredlight}).

\citet{2013AAAcke} claim that the continuum opacity source in the circumbinary disk of HR4049 can be identified as amorphous carbon, 
by confronting radiative transfer models with near- and mid-IR interferometric data and the IR ISO spectrum. They state 
that the identification can be made in essence because the temperatures reached by various dust species at a given physical 
distance to the star are very different. As an explanation for the origin of the carbonaceous material, they propose that ongoing mass 
loss from the depleted photosphere of the post-AGB primary consists of carbon-bearing molecules. This is consistent with 
various observations \citep[][]{1989ApJGeballe,2012MNRASRoberts,2013AAAcke,2014ApJMalek}. According to \citet{2013AAAcke},
this material may then form a veil of carbon-rich dust over the pre-existing silicate-rich disk.
It remains to be demonstrated, however, that this outflowing material influences the structure and appearance of the disk, 
such that this is detectable.

The models used by \citet{2013AAAcke} were rather simple and broke down beyond 10~$\mu$m, 
as the dust consisted of only 0.01~$\mu$m grains of a single species. Although the effect of very small metallic Fe 
vs. amorphous C grains is very different in terms of heating, the influence of the larger grains, which constitute the bulk of the 
dust mass, on the temperature structure cannot be neglected \citep[see also][]{2012AAMulders}. Our results show that their conclusion about the 
composition of the dust component in the HR4049 disk needs to be carefully reevaluated. Our best Fe and C 
models give very similar fits to our observations of 89 Her; and the latter constraints are similar 
to those relied upon by \citet{2013AAAcke}. As Figs.~\ref{figure:bestModelTStructure} and~\ref{figure:bestModelDensityStructure} show,
the inner temperature structures are not very different between our two best models, while the density structures are different. 
We stress that the disk structure needs to be taken into account, including the important effect of dust settling, like was done in this work. 

No evidence exists for the presence of carbon-rich gas nor dust in the 89 Herculis system. A large-scale outflow has been detected 
in CO \citep{2007AABujarrabal}, but its origin within the central object is unknown (see Sect.~\ref{section:outflow}). 
Also, the photospheric depletion pattern in 89 Her is not very pronounced \citep{2011BaltAKipper}, in contrast to HR4049, which 
is one of the most depleted objects \citep{1991AAWaelkens} among the known sample of Galactic post-AGB binaries. Hence, from the perspective of 
abundances, the photosphere of 89 Her contains sufficient Si and Fe to form O-bearing dust in the outflow. While our 
best-fitting C model matches the observations mildly better than our best-fitting Fe model, it seems much harder to explain the formation of
amorphous carbon dust in the 89 Her system than in HR4049. This is the case, in particular, for a homogeneous dust composition throughout the disk, 
as is assumed here.

\subsection{The scattered light} \label{section:scatteredlight}
Finally, we discuss the main result of this paper, which is the inability of our disk models to reproduce the detected optical scattered light.
Despite the extensive grid of calculated models, and the two dust compositions that were explored, not a single model produces sufficient scattered 
light at optical wavelengths. Moreover, those models that approach the required optical and resolved flux fraction do so at the cost of 
a too large scattering surface area and/or too high near-IR fluxes.

Considering the single-scattering albedos in Fig.~\ref{figure:albedos}, it becomes clear that another change in dust composition will not 
increase the scattered light by much. Doubling the iron abundance to 30\% in mass fraction (an extreme situation)
only slightly affects the albedo. One might expect that removing the smallest grains ($\leq$0.03~$\mu$m) from the population, 
and including more grains in the $\sim$0.03--0.3~$\mu$m size range by adapting the grain size power-law index, might increase
the amount of scattered light. In contrast to the former, the latter grains have a high albedo in the 
optical wavelength range, and do not yet have a scattering phase function that is strongly forward-peaked. 
\citet{2013AAMulders} showed that this strong forward peak in the scattering phase function of grains $\geq$0.3~$\mu$m is the cause 
for the faintness of protoplanetary disks in optical scattered light. We therefore recomputed our best-fit model, now excluding the 
smallest grains, and found no noticeable difference in the temperature structure nor in the predicted observables of our model. 
Although the smallest grains have high extinction coefficients, their density (see Fig.~\ref{figure:bestModelDensityStructure})
is very low and they thus do not dominate the opacity in this model. 

The small deficiencies of our best-fit Fe model with respect to the SED, mentioned in Sect.~\ref{subsect:nearIR}, point to a lack of small grains, and 
hence a power-law index that has too shallow a grain size distribution. Increasing the number of small grains is, however, 
prohibited by the constraints set in the H band. Our best amorphous carbon model partly resolves some of these issues, 
but still overestimates the H band flux and underpredicts the far-IR flux. Moreover, 
the origin of the carbon would be difficult to explain in this system. Finally, the visibilities are not perfectly fitted by either model,
as an even flatter slope seems to be required at the intermediate-to-long spatial frequencies.

We conclude that no reasonable disk model can match \textit{all} of the observations, because two distinct geometric structures 
are needed to accommodate the near-IR and optical constraints simultaneously. The structure of our best models is too flat
at the inner rim to scatter a sufficient amount of the optical stellar flux at this radius into our line of sight, which is 
needed to fit our optical observations. In contrast, this flattened inner rim is needed to fit the near-IR data, and in particular the visibilities. 
So to explain the deficiencies, we propose that the basic geometry needs to be adapted: an outflow in optical scattered light 
should be included in the model of the system in addition to the circumbinary disk. As discussed in Paper I, there 
is some indirect evidence from polarimetric data \citep{1987AAJoshi} that 89 Her may also have a low-density 
component of small dust grains that can be associated with a bipolar structure. 

\subsection{The global picture}\label{section:outflow}
Several questions remain about our interpretation of this system. The CO hourglass outflow is detected on a very different scale 
\citep[$\sim$10$\arcsec$,][]{2007AABujarrabal} than the scales traced by our interferometric observations. Moreover, the same 
authors estimated the lifetime of this hourglass at $\sim$3500~yr. During this time the dynamics of the inner system may have changed. 
The H$\alpha$ line shows a clear P Cygni type of profile \citep{1993AAWaters}, but the properties of the current outflow 
may be very different (e.g., more jetlike) from the CO hourglass. Based on our current modeling of the existing observations, we cannot conclude 
whether the optically scattering dust is dynamically linked to the hourglass, nor to a speculative more recent jetlike outflow.

The driving mechanism of the CO hourglass is unknown, so there are no independent constraints on the density structure nor on the grain 
properties of a potential dust component in it. Direct dust condensation in a wind seems unlikely given the effective temperature (6550~K) of the star, 
but dust grains from the circumbinary disk may be trapped in an accretion flow that is subsequently blown away in the polar directions. 
The launching mechanism of the gas and dust (i.e., the exact path followed by either) remains an open question in this framework
\citep[see, e.g., the model calculations of][]{2012ApJBans}.
The smoothness of the inner disk rim as found in this work, may be an indication for such a scenario where ongoing accretion 
is linked to the formation of the outflow. More post-AGB disks should be studied with near-IR interferometry to investigate 
a possible link between the inner disk shape and the presence of an outflow in the system.

Additional quantitative models that include the disk as well as a dust component in the (hourglass) outflow are 
needed in combination with more constraining interferometric observations.
The parameter values of our best-fit disk models may be affected by the presence of dust in the outflow. 
It cannot be excluded that the outflow also contributes near- and/or mid-IR flux, which was neglected in the current framework. 
It might even be expected that this additional structure of low optical depth in/above the system presents a natural way of smoothing 
the intensity profile when viewed face-on. With this additional structural component, however, many new free parameters 
would be introduced in the model, in addition to the disk parameter space that needs to be reevaluated simultaneously. 
Therefore, degeneracies can be expected that the current data are incapable of resolving. Given the correspondence between 
the SED and the albedo of 0.1~$\mu$m grains (see Fig.~\ref{figure:albedos}), we nevertheless predict that an outflow 
consisting of such grains, and covering a large volume but with a low optical depth, might give rise to a significant 
flux of optical scattered light without producing too much IR thermal emission. It is, however, beyond the scope of this paper to explore such models.




\section{Conclusions} \label{sect:conclusions}
In this paper we have computed an extensive grid of radiative transfer models of passive disks in hydrostatic equilibrium, using a grain size distribution 
that includes large grains and a consistent treatment of grain-size-dependent dust settling. Such models very successfully match the energetics and 
spatial distribution of the circumstellar material in the 89 Herculis system, as probed by the full IR SED and our IR interferometric observables.
We have extended the SED with precise SPIRE flux measurements in the 200-600~$\mu$m wavelength range, thereby putting strong constraints on the 
size distribution of the dust grains. Our results confirm that the binary in 89 Herculis is surrounded by a compact 
and stable circumbinary disk in Keplerian rotation, wherein large dust grains form and settle to the midplane as in our best models.
We also showed that no firm conclusions can be drawn on the composition of the dust, based on the observables used in this paper, without taking 
the full disk structure into account.

In the optical, however, only about half of the required circumstellar scattered flux is predicted by our disk models, 
albeit at the correct angular size scale. Although the parameter space is vast, and not all avenues in terms of disk models 
have been investigated in full detail \citep[see, e.g.,][where the possible role of magnetic fields is investigated in the case of disks around 
Herbig stars]{2014ApJTurner}, our results show that to resolve the remaining discrepancies, an additional component with some scattering material in 
our direct line-of-sight to the central binary is probably needed in the 89 Her system. The very smooth 
emission profile at near-IR and the compactness at optical wavelengths are difficult to reconcile within a geometrical framework 
that only consists of a puffed-up circumbinary disk in hydrostatic equilibrium. 

Whether the dusty material is in the (detected molecular) bipolar outflow, in a halo near the inner disk rim, or in yet another structure,
is unclear at the moment. \citet{2011AAKrijt} suggested that halos could form near the inner edge of protoplanetary disks by 
dynamical interactions and collisions between planetesimals. In light of our speculative detection of a ``gap'' in the MIDI data 
of 89 Her, such a scenario cannot be excluded a priori. It seems more likely, however, that the dust is present in the bipolar outflow, 
given that dust is detected in a variety of outflows around other post-AGB objects. The central question that remains relates to 
the origin of the outflow and its dust component. The grains in the outflow need to be $\geq$0.1~$\mu$m to efficiently 
scatter up to wavelengths of $\sim$1~$\mu$m. The details of 
dust condensation in the close neighborhood of oxygen-rich (post-)AGB stars are not well understood, so the origin of these grains 
remains to be demonstrated. The mass reservoir from which the gas component in the outflow originates is not firmly identified either. 
The mass could either be lost by the central star(s) or by the inner disk. 
More (precise) observations are required to make an exploration of these complex model geometries fruitful.

We stress that our best-fit models of the inner geometry as listed in Table~\ref{table:MCMaxInput} 
were determined under the assumption that there is only a binary with a circumbinary disk in the system! On the other hand, 
we also underline that even though the outflow likely contributes scattered light at optical wavelengths, the difference 
in size distribution between the grains in the disk and in the outflow means that the (far-)IR emission is definitely dominated by the disk.

The complex inner geometry, the large scattering component, and the unknown physical connection between the CO hourglass outflow 
at large scales and the inner disk, indicates that multi-wavelength interferometric imaging is likely needed to constrain 
a global model of the 89 Her system. More clues on the 3D position of the scattering material can also come from stronger observational constraints 
on the wavelength dependence of the optical scattering, and from spatially resolving the system in different polarization states, 
since the various geometries can be expected to result in different polarimetric signatures.



\begin{acknowledgements}
This research has made use of the SIMBAD database, operated at CDS, Strasbourg, France. This research has made use of NASA's Astrophysics Data System. 
The Palomar Testbed Interferometer was operated by the NASA Exoplanet Science Institute 
and the PTI collaboration. It was developed by the Jet Propoulsion Laboratory, California Institute of Technology with funding 
provided from the National Aeronautics and Space Administration. We acknowledge with thanks the variable star observations 
from the AAVSO International Database contributed by observers worldwide and used in this research. We also thank B. Acke for his input 
and the useful discussions about this work. Finally, we are grateful for the constructive comments and questions provided by the referee.
\end{acknowledgements}

\bibliographystyle{aa}
\bibliography{aa_89Her2}

\end{document}